\documentclass[pdftex,twocolumn,epjc3]{svjour3}          
\usepackage{lipsum}
\newcommand{\mnras}{MNRAS}
\newcommand{\aap}{A\& A}
\newcommand{\apjl}{ApJL}
\newcommand{\apj}{ApJ}
\newcommand{\prd}{Phys. Rev. D.}
\newcommand{\prl}{Phys. Rev. Lett.}
\newcommand{\nat}{Nature}
\newcommand{\physrep}{PhysRep}
\usepackage{romannum}
\usepackage[mathlines]{lineno}
\usepackage{graphicx}
\usepackage{dcolumn}
\usepackage{bm}
\usepackage{amsmath,amssymb}

\smartqed  
\RequirePackage{fix-cm}
\RequirePackage{graphicx}
\RequirePackage{mathptmx}      
\RequirePackage{flushend}
\RequirePackage[numbers,sort&compress]{natbib}
\RequirePackage[colorlinks,citecolor=blue,urlcolor=blue,linkcolor=blue]{hyperref}

\journalname{Eur. Phys. J. C}

\begin{document}\sloppy
\title{The structure of the ultrarelativistic prompt emission phase and the properties of the black hole in GRB 180720B}


\author{F.~Rastegarnia\thanksref{e1,addr1,addr2,addr3,addr4}
        \and
        R.~Moradi\thanksref{e2,addr1,addr2,addr5} 
        \and
        J.~A.~Rueda\thanksref{e3,addr1,addr2,addr3,addr4,addr6} 
        \and
        R.~Ruffini\thanksref{e4,addr1,addr2,addr7} 
        \and
        Liang~Li\thanksref{e5,addr1,addr2,addr7} 
        \and
        S.~Eslamzadeh\thanksref{addr1,addr2,addr3,addr4} 
        \and
        Y.~Wang\thanksref{e7,addr1,addr2,addr5} 
        \and
        S.~S.~Xue\thanksref{addr1,addr2} 
}

\thankstext{e1}{e-mail: fatemeh.rastegarnia@edu.unife.it}
\thankstext{e2}{e-mail: rahim.moradi@inaf.it}
\thankstext{e3}{e-mail: jorge.rueda@icra.it}
\thankstext{e4}{e-mail: ruffini@icra.it}
\thankstext{e5}{e-mail: liang.li@icranet.org}
\thankstext{e7}{e-mail: yu.wang@uniroma1.it}

\institute{ICRA, Dipartimento di Fisica, Sapienza Universit\`a  di Roma, Piazzale Aldo Moro 5, I--00185 Roma, Italy\label{addr1}
          \and
          International Center for Relativistic Astrophysics Network, Piazza della Repubblica 10, I--65122 Pescara, Italy\label{addr2}
          \and
          ICRANet-Ferrara, Dipartimento di Fisica e Scienze della Terra, Universit\`a degli Studi di Ferrara, Via Saragat 1, I--44122 Ferrara, Italy\label{addr3}
          \and
          Dipartimento di Fisica e Scienze della Terra, Universit\`a degli Studi di Ferrara, Via Saragat 1, I--44122 Ferrara, Italy\label{addr4}
          \and
          INAF -- Osservatorio Astronomico d'Abruzzo,Via M. Maggini snc, I-64100, Teramo, Italy\label{addr5}
          \and 
          INAF, Istituto di Astrofisica e Planetologia Spaziali, Via Fosso del Cavaliere 100, 00133 Rome, Italy\label{addr6}
          \and
          INAF, Viale del Parco Mellini 84, 00136 Rome, Italy\label{addr7}
}

\date{Received: date / Accepted: date}

\maketitle

\begin{abstract}
In analogy with GRB 190114C, we here analyze the ultrarelativistic prompt emission (UPE) of GRB 180720B observed in the rest-frame time interval $t_{\rm rf}=4.84$--$10.89$~s by Fermi-GBM. We reveal the UPE hierarchical structure from the time-resolved spectral analysis performed in time sub-intervals: the spectrum in each shorter time interval is always fitted by a composite blackbody plus cutoff power-law model. We explain this structure with the \textit{inner engine} of binary-driven hypernova (BdHN) model operating in a quantum electrodynamics (QED) regime. In this regime, the electric field induced by the gravitomagnetic interaction of the newborn Kerr BH with the surrounding magnetic field is overcritical, i.e., $|{\bf E}|\geq E_c$, where $E_c=m_e^2 c^3/(e\hbar)$. The overcritical field polarizes the vacuum leading to an $e^+~e^-$ pair plasma that loads baryons from the surroundings during its expansion. We calculate the dynamics of the self-acceleration of the pair-electromagnetic-baryon (PEMB) pulses to their point of transparency. We characterize the quantum vacuum polarization process in the sequences of decreasing time bins of the UPE by determining the radiation timescale, Lorentz factors, and transparency radius of the PEMB pulses. We also estimate the strength of the surrounding magnetic field $\sim 10^{14}$ G, and obtain a lower limit to the BH mass, $M=2.4~M_\odot$, and correspondingly an upper limit to the spin, $\alpha = 0.6$, from the conditions that the UPE is powered by the Kerr BH extractable energy and its mass is bound from below by the NS critical mass.
\end{abstract}

\section{Introduction}\label{sec:1}

\textcolor{black}{The binary-driven hypernova (BdHN) model has been proposed for the description of long-duration gamma-ray bursts (GRBs), following the induced gravitational collapse (IGC) scenario \cite{2012ApJ...758L...7R}. The progenitor is a binary system composed of a carbon-oxygen core (CO$_{\rm core}$) and a neutron star (NS) companion. These CO$_{\rm core}$-NS binaries are expected to form in the late stages of the binary evolution of massive binaries, e.g., $\sim 15+12 M_\odot$ zero-age main-sequence (ZAMS) stars \cite{2020ApJ...893..148R}. The more massive star undergoes core-collapse supernova (SN) and creates a NS when its thermonuclear evolution is over. After multiple common-envelope phases and binary interactions during the X-ray binary phase of the system (see, \cite{2014ApJ...793L..36F, 2015PhRvL.115w1102F}, and references therein), the hydrogen and helium envelopes of the less massive main-sequence star are stripped, leading to the formation of the CO$_{\rm core}$  (see \cite{2015PhRvL.115w1102F}, and  Fig.~1 and section 2 in \cite{2020ApJ...893..148R}). The system at this point is a CO$_{\rm core}$-NS binary in tight orbit, which is taken as the initial configuration of the BdHN model for long GRBs \cite{2014ApJ...793L..36F, 2015ApJ...812..100B,2016ApJ...833..107B,2019ApJ...871...14B,2020ApJ...893..148R}.}


In the last decade, theoretical progress in the analysis of BdHNe, including three-dimensional numerical simulations, has identified their sequence of physical events \cite{2012ApJ...758L...7R, 2012A&A...548L...5I, 2014ApJ...793L..36F, 2015PhRvL.115w1102F, 2015ApJ...812..100B, 2016ApJ...833..107B, 2019ApJ...871...14B}. The gravitational core-collapse event of the CO$_{\rm core}$ forms a newborn NS ($\nu$NS) at its center and powers a \textcolor{black}{type Ic} SN explosion. The SN ejecta is partially accreted by the $\nu$NS because of matter fallback and partially by the NS companion at hypercritical (i.e., highly super-Eddington) rates due to a powerful neutrino emission process occurring on top the NS surface \cite{2016ApJ...833..107B, 2018ApJ...852..120B}. 
\textcolor{black}{The fallback accretion onto the $\nu$NS  contribute to the energy of prompt emission and spins up the $\nu$NS (Becerra et al., submitted; see also \cite{2019ApJ...871...14B} and Yu et al., in press). The $\nu$NS rotational energy powers the synchrotron emission leading to the afterglow \cite{2012ApJ...758L...7R, 2014ApJ...793L..36F, 2019ApJ...871...14B, 2021MNRAS.504.5301R, 2021IJMPD..3030007R}.} For orbital periods of a few minutes, the NS companion reaches the critical mass for gravitational collapse leading to the formation of a rotating (Kerr) black hole (BH). These systems have been called BdHN I, \textcolor{black}{which explain the subclass of energetic long GRBs with $E_{\rm iso} \gtrsim  10^{52}$ erg. Up to now, 380 long GRBs have been interpreted as BdHNe I \cite{2021MNRAS.504.5301R}}. \textcolor{black}{Therefore, three pillars of BdHN I, responsible for different episodes of this subclass of long GRBs are: 1) SN, 2) BH, and 3) $\nu$NS. The interplay between  these three components leads to different episodes of BdHN I.} For longer orbital periods, the NS companion does not reach the critical mass and hold stable as a more massive, fast rotating NS. These systems have been called BdHN II, \textcolor{black}{which explain the subclass of energetic long GRBs with $E_{\rm iso} \lesssim  10^{52}$ erg}. \textcolor{black}{The emergence of the optical SN naturally expected in the BdHN scenario (see e.g., \cite{2019GCN.25657....1P, 2019GCN.23715....1R, 2019GCN.25677....1D, 2013GCN.14526....1R}) completes the BdHN approach.}

The experimental verification of the entire sequence of Episodes in a BdHN I has been recently achieved in GRB 190114C \citep{2021PhRvD.104f3043M} \textcolor{black}{and GRB 180720B \citep{2021arXiv210309158M} }. Thanks to the high quality of the data and the brightness of \textcolor{black}{these sources}, we have identified through a detailed time-resolved analysis the following Episodes of a BdHN I \cite{2021arXiv210309158M, 2021PhRvD.104f3043M}: the emission from the $\nu$NS (the $\nu$NS-rise); the BH formation, \textcolor{black}{known as BH-rise,} originating the ultrarelativistic prompt emission (UPE) phase; the formation of the \textit{cavity} around the newborn BH, \textcolor{black}{formed in the gravitational collapse of the companion NS to the BH, and further depleted by the UPE phase} \cite{2019ApJ...883..191R}; \textcolor{black}{the soft and hard X-ray flares (SXF and HXF) originating from the interaction of the UPE phase with the expanding SN ejecta \citep{2018ApJ...869..151R}};  the X-ray afterglow powered by the rapidly rotating $\nu$NS \cite{2018ApJ...869..101R, 2019ApJ...874...39W, 2020ApJ...893..148R}, and the gigaelectronvolt (GeV) emission from the BH following the UPE \cite{2019ApJ...886...82R, 2021A&A...649A..75M}. \textcolor{black}{We discuss in section~\ref{sec:observation} the observational identification of the above episodes in the case of GRB 180720B.}

In this article, \textcolor{black}{we perform a time-resolved spectral analysis of the UPE phase of GRB 180720B and interpret it in the context of the BH formation in a BdHN I. A most relevant result of this kind of analysis has been the discovery of the hierarchical structure of the UPE phase of GRB 190114C.} The spectrum on rebinned shorter time intervals shows always a similar blackbody plus cutoff power-law (BB + CPL) model during the entire UPE phase \cite{2021PhRvD.104f3043M}. The explanation of such a hierarchical structure of the UPE phase has been found in the sequence of microphysical elementary events, in the quantum electrodynamics (QED) regime of vacuum polarization, that occurs in the formation of the BH in the inner engine of the GRB \cite{2019ApJ...886...82R}. The inner engine is the system composed of the newborn rotating BH surrounded by a uniform magnetic field, aligned with the BH rotation axis, and the low-density ($\sim 10^{-14}$ g cm$^{-3}$) matter of the SN ejecta in the cavity.

The physical process, which combines the pure general relativistic effect of \textit{gravitomagnetism} and QED works as follows. The gravitomagnetic interaction of the Kerr BH with the magnetic field induces an electric field, and the structure of such an electromagnetic field has been modeled with the Papapetrou-Wald solution \cite{1974PhRvD..10.1680W}. The intensity of the induced electric field is proportional to the BH spin parameter and the magnetic field strength.
\textcolor{black}{The newborn BH is not charged. The interaction of the gravitomagnetic field of the Kerr BH with the surrounding magnetic field, $B_0$, induces an electric field around the BH. This electric field is nearly radial, and despite its quadrupolar nature, decreases with distance roughly as $1/r^2$. Hence, it is possible to define an ``effective charge'', $Q_{\rm eff}$, as the proportionality constant of such a field, i.e., $E \approx Q_{\rm eff}/r^2$, where \citep{2019ApJ...886...82R, 2020EPJC...80..300R, 2021A&A...649A..75M}
\begin{eqnarray}\label{eq:EFCH}
  Q_{\rm eff}=\frac{G}{c^3}2 B_0 J.
\end{eqnarray}}

\textcolor{black}{It can be shown that the Papapetrou-Wald solution, due to theorems by \citet{PhysRevLett.27.529}, \citet{Waldthesis} and \citet{DeWitt:1973uma}, produces the unique solution which, at perturbative level, represents the transformation of the Kerr BH into a charged rotating Kerr-Newman BH, with effective charge given by the above equation, i.e., Eq.~(\ref{eq:EFCH}) of the paper. It can be \textcolor{black}{indeed} checked that for relatively low values of the spin parameter $\alpha = c J/(G M^2) \lesssim 0.6$, the Papapetrou-Wald solution can be approximated by the Kerr-Newman solution \citep{2021A&A...649A..75M}. We take advantage of the above property to estimate the energy and the QED effect with the Kerr-Newman geometry for which an analytic expression for the energy of the dyadoregion has been derived in \citet{2009PhRvD..79l4002C}. In fact, the difficulty of the origin of a charged BH is overcome by the idea of the effective charge.}

 The inner engine in the UPE phase operates in an overcritical QED regime in which the induced electric field is larger than the critical field for vacuum polarization, i.e., $E > E_c$, where
\begin{equation}\label{eq:Ec}
    E_c = \frac{m_e^2 c^3}{e\hbar} \approx 1.32\times 10^{16}\,\,\rm V/cm.
\end{equation}
The MeV radiation of the UPE and its associated hierarchical structure is explained by the inner engine in this overcritical regime, and is powered by the rotational energy of the Kerr BH. In this article, we focus on the UPE phase of GRB 180720B. 

The overcritical electric field \textcolor{black}{of the inner engine} generates an initially optically thick $e^+e^-$ plasma that self-accelerates under its own internal pressure while engulf ambient baryons. The first numerical simulations of the expanding optically thick pair electromagnetic-baryon plasma, called \textit{PEMB} pulse, were presented in \cite{1999A&A...350..334R}. \textcolor{black}{For instance, for a BH mass $10 M_\odot$ and effective charge to mass ratio of $\sim 0.1$, adopted for GRB 991216 \citep{2004IJMPD..13..843R}, the produced $e^+e^-$ plasma pairs lie between the radii $r_1=6 \times 10^6$ cm and $r_2=2.3 \times 10^8$ cm, with total energy of $E_{\rm tot}= 4.8 \times 10^{53}$ erg, and the total number of pairs is $N_{e^+e^-}=2 \times 10^{58}$. This leads to the  pair number density of $10^{32} \lesssim \overline{n}_{e^+e^-} \lesssim 10^{37}$  cm$^{-3}$ and the optical depth of $\tau \sim \sigma_T~ \overline{n}_{e^+e^-} \times [r_2-r_1] \sim  10^{16}$--$10^{21} \gg 1$, being $\rm \sigma_T=6.6 \times 10^{-25}$~cm$^{-2}$ the Thomson cross-section. Such an optically thick PEMB pulse self-accelerates outward reaching ultrarelativistic velocities \citep{1999A&A...350..334R, RSWX2} up to Lorentz factors of $\Gamma \sim 300$ at transparency and emits MeV photons. The observation of such thermal photons signs the first evidence of the Kerr BH formation, i.e., the BH-rise. } The high Lorentz factor guarantees the avoidance of the so-called GRB compactness problem \cite{1975NYASA.262..164R,2004RvMP...76.1143P}. 

\textcolor{black}{In \citet{2020ApJ...893..148R} (see Section 7 therein), it has been discussed that numerical simulations of BdHN I point to the possible presence of a torus-like distribution of matter with higher density on the equatorial plane that can serve to anchor the magnetic field. The physical process leading to the UPE phase requires the presence of low-density ionized matter on the polar regions, i.e., above and below the BH. Therefore, the presence of matter with higher density on the equatorial plane, providing it is not as massive as to change the assumed spacetime geometry, does not interfere with the production of the pair plasma around the BH.}

The emitted energy \textcolor{black}{in the UPE phase} is paid by the rotational energy of the BH, hence it reduces its angular momentum, and consequently the intensity of the induced electric field. This process continues until the electric field reaches the value $E_c$ and no more pairs can be created via vacuum breakdown. We here analyze all the above process occurring during the UPE phase in the case of GRB 180720B, which is another BdHN I and its data quality allows us to perform a detailed time-resolved spectral analysis analogous to the one applied successfully to GRB 190114C in \cite{2021PhRvD.104f3043M}. We confirm in this paper the presence in the $10$ keV--$10$ MeV energy band the very same hierarchical structure of the UPE phase in GRB 180720B already found in GRB 190114C. We simulate the above physical process of the inner engine that explains the UPE energetics, luminosity and spectrum and infer from it the magnetic field strength, the initial mass and spin of the BH, and their time evolution.

The electro-vacuum Papapetrou-Wald solution used in the inner engine differs from other models of the electromagnetic field structure around astrophysical BHs (see, e.g.,  \cite{2005MNRAS.359..801K, 2019PhRvL.122c5101P}). A detailed theoretical review of such models is presented by \citet{2005MNRAS.359..801K}. In those models, the field structure and parameters enforce the condition $\mathbf{E} \cdot \mathbf{B} \neq 0$, while in the Papapetrou-Wald solution naturally exist regions where such a condition is naturally satisfied in the Kerr BH surroundings. In the BdHN I, the BH is surrounded by a very-low dense plasma in which the screening of the electric field is unlikely to occur, guaranteeing the stability of the Papapetrou-Wald electromagnetic field structure. This differs from the environment envisaged for describing the mechanism of generating powerful relativistic jets from a black hole system in AGNs and x-ray binary systems, e.g., in \citet{2005MNRAS.359..801K} and \citet{2019PhRvL.122c5101P}, where the surrounding plasma has a much larger density and may cause the screening of the electric field. Therefore, the inner engine parameters and physical processes are different with respect to these models and have been guided by the GRB analysis. Specifically, our approach gives a theoretical explanation to the time-resolved spectral analysis of the UPE phase, and the MeV luminosity observed by Fermi-GBM.

{The article is organized as follows.} In Sec.~\ref{sec:observation}, we present the observations of GRB 180720B by different satellites and then introduce the 6 Episodes of this GRB obtained from the observations. In Sec.~\ref{sec:heir}, we perform the time-resolved spectral analysis during the UPE phase of GRB 180720B {thanks to} the recent progress in the understanding of the UPE phase of GRB 190114C \cite{PhysRevD.104.063043}, and the high signal-to-noise (S/N) ratio of the Fermi-GBM data during the UPE phase of GRB 180720B. 
In Sec.~\ref{sec:5}, {we introduce the structure of the electromagnetic field used to study} the properties of inner engine of GRB 180720B. {We also discuss the physical differences of this electromagnetic field and the operation of the inner engine to extract the rotational energy of the BH with the existing literature on the subject.} In Sec.~\ref{sec:massupe}, we determine the mass and spin of the BH in the inner engine of GRB 180720B during the UPE phase {and in the subsequent evolution}. In Sec.~\ref{sec:vac}, {we describe the vacuum polarization process in the inner engine and how it originates the UPE phase.} In Sec.~\ref{sec:12}, {we analyze} the compactness problem and the general formulation of transparency condition during the UPE phase. In Sec.~\ref{sec:magnetic}, we obtain the magnetic field and BH parameters at transparency point during the UPE phase. We follow the quantum vacuum polarization process down to a timescale of $\tau \sim 10^{-9}$~s, marking the hierarchical structure of the UPE imposed by the observed luminosity and the electromagnetic configuration of the inner engine during the UPE phase. We compute the value of the magnetic field, $B_0$,  the Lorentz factors, the baryon loads, the energy emitted and radii at the transparency of each PEMB pulse. {We also discuss relevant differences between our approach and different models in the literature, e.g. by \citet{2005MNRAS.359..801K} and \citet{2019PhRvL.122c5101P}.} In Sec.~\ref{sec:conc}, we {summarize the conclusions} of this work. 

\begin{figure*}
    \centering
\includegraphics[width=19 cm]{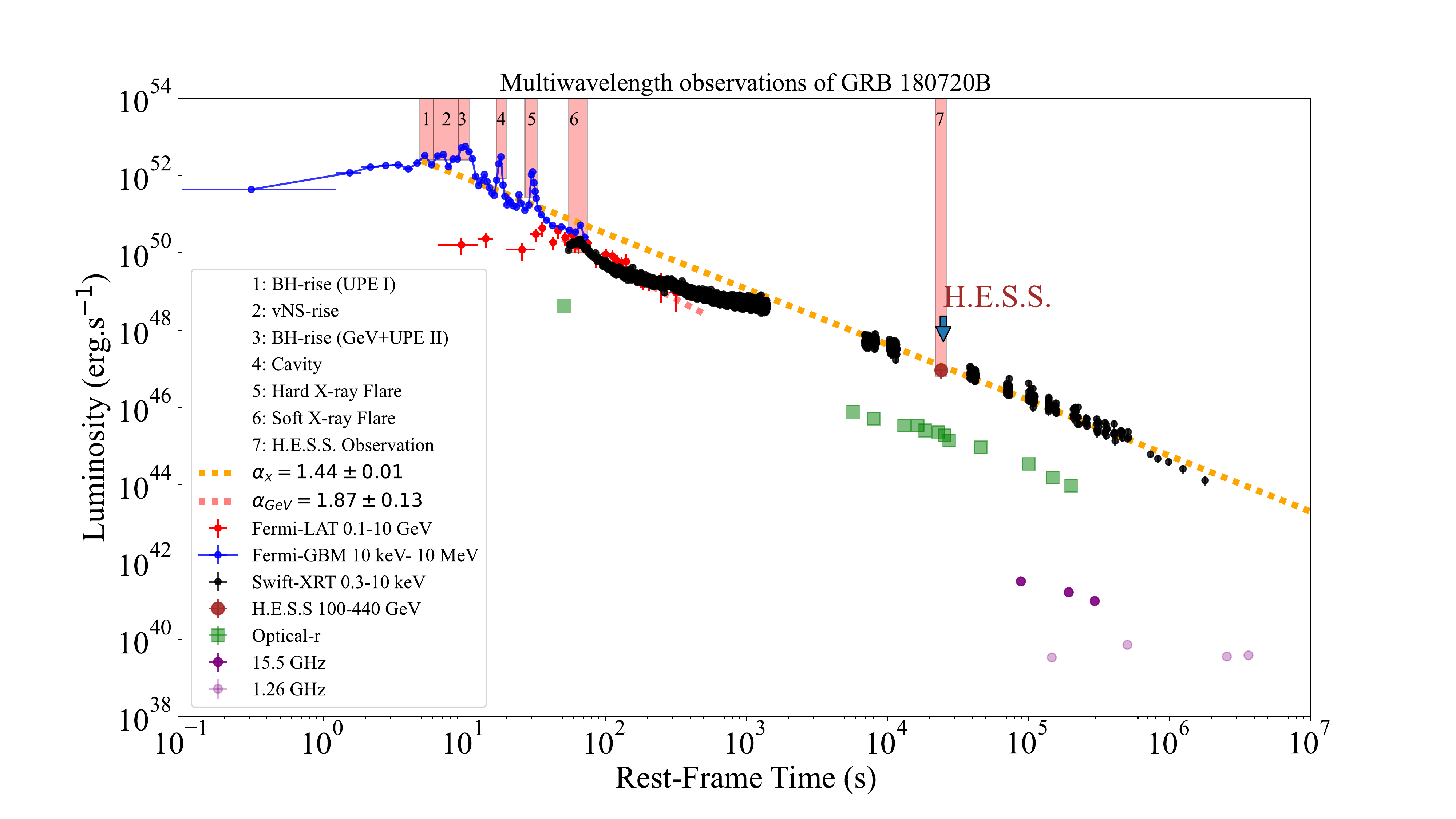}
     \caption{   Luminosity of GRB 180720B in the cosmological rest-frame of the source. Blue circles: obtained from \textit{Fermi}-GBM in the $10$~keV--$10$~MeV energy band. Red circles:  obtained from \textit{Fermi}-LAT in the $0.1$--$10$~GeV energy band. The Episode 1 shows the {BH
     }--rise {(UPE I)} from $~t_{\rm rf}=4.84$~s to $~t_{\rm rf}=6.05$~s. The Episode 2 shows the $\nu$NS--rise from $~t_{\rm rf}=6,05$~s to $~t_{\rm rf}=9.07$~s. The Episode 3 represents the {second} UPE phase {(UPE II)} from $~t_{\rm rf}=9.07$~s to $~t_{\rm rf}=10.89$~s. The Episode 4 shows the Fermi-GBM the \textit{cavity} from $t_{\rm rf}=16.94$~s to $~t_{\rm rf}=19.96$~s. The Episode 5 the hard X-ray Flares (HXF), from $t_{\rm rf}= 28.95$~s to $t_{\rm rf}= 34.98$~s. The Episode 6, the soft X-ray Flares (SXF), from $t_{\rm rf}= 55$~s to $t_{\rm rf}= 75$~s. The rest-frame $0.1$--$100$~GeV luminosity light-curve of GRB 180720B after UPE phase is best fitted by a power-law with slope of $\alpha_X=1.87\pm 0.13$ and, amplitude of $(4.6\pm 2.9)\times 10^{53}$~erg~s$^{-1}$. The purple circles present the radio data from AMI--LA retrieved from \citet{2020MNRAS.496.3326R}. The green squares represent the r-band optical data retrieved from \citet{2019Natur.575..464A}. }\label{fig:data}%
\end{figure*}

\section{Observational data of GRB 180720B} 
\label{sec:observation}

On July 20, 2018, GRB 180720B triggered the Fermi-GBM at 14:21:39.65 UT \citep{GCN22981}, the CALET Gamma-ray Burst Monitor at 14:21:40.95 UT \citep{GCN23042}, the Swift-BAT at 14:21:44 UT \citep{GCN22973}, the Fermi-LAT at 14:21:44.55 UT \citep{GCN22980}, and the Konus-Wind at 14:21:45.26 UT \citep{GCN23011}. The X-ray telescope (XRT) on board the Neil Gehrels Swift Observatory (hereafter Swift), began observing $91$ s after the Fermi-GBM trigger \citep{GCN22973}, MAXI/GSC started at $296$ s \citep{GCN22993} and NuStar covered later times from $243$ ks to $318$ ks \citep{GCN23041}. Just $78$ s after the Fermi-GBM trigger, the $1.5$-m Kanata telescope performed optical and NIR imaging polarimetry of the source field and found a bright optical R-band counterpart within the the Swift-XRT error circle, observed by HOWPol and HONIR attached to the $1.5$-m Kanata telescope \citep{GCN22977}. Additional observations followed by optical, infrared and radio telescopes \citep{GCN22976,2018GCN.22977....1S,GCN22983,GCN22985,GCN22988,GCN23017,GCN23020,GCN23021,GCN23023,GCN23024,GCN23033,GCN23037,GCN23040,2019Natur.575..464A}. 

Following the optical observation, redshift z = 0.654 was identified from the Fe II and Ni II lines by the VLT/X-shooter telescope \citep{GCN22996}. This allows to determine the cosmological rest frame time $t_{\rm rf}=t_{\rm obs}/(1+z)$, being $t_{\rm obs}$ the observed time, as well as the isotropic energy ($E_{\rm iso}$) and the isotropic luminosity ($L_{\rm iso}$) of this source. GRB 180720B has isotropic energy of $E_{\rm iso}=5.92 \times 10^{53}$~erg during the $T_{90}$ of the Fermi-GBM. The sub-TeV ($100$--$440$~GeV) observation by the High Energy Stereoscopic System (H.E.S.S.) has been also reported for this GRB \citep{2019Natur.575..464A}. The diversity and the statistical significance of the observed data have made this GRB one of the proper candidates to test the GRB models. The luminosity light-curve in radio, optical, and gamma-rays of the GRB 180720B is shown in Fig.~\ref{fig:data}.

\subsection{The Episodes of GRB 180720B}

Six different episodes relating to six different astrophysical processes have been recently identified in the time domain analysis of GRB 180720B \citep{2021arXiv210309158M}: 

\begin{figure*}
\centering
\includegraphics[width=0.99\hsize,clip]{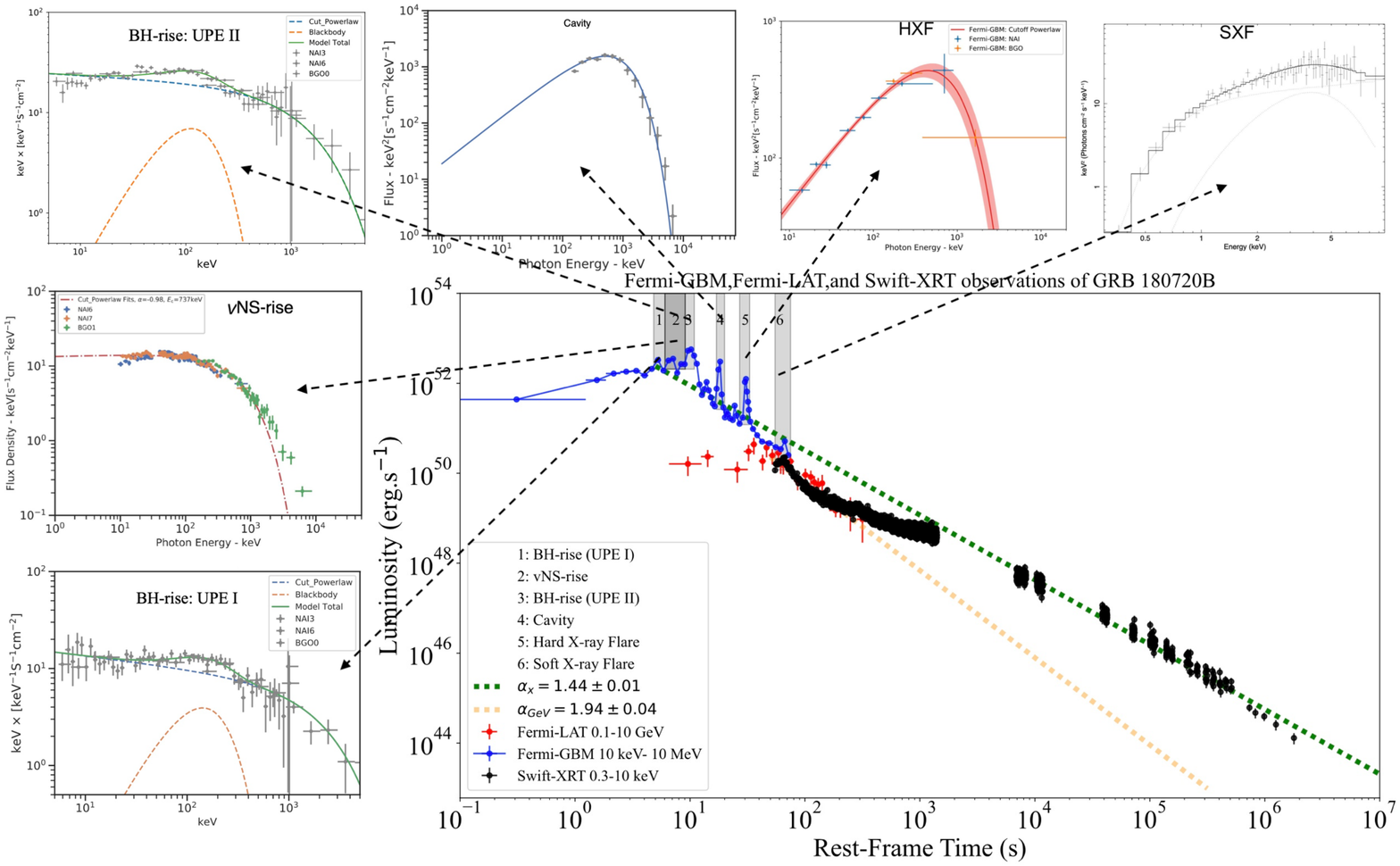}
\caption{\textcolor{black}{Luminosity light-curves and spectra related to the different Episodes identified in GRB 180720B. Plots and best fits are reproduced from \citep{2021arXiv210309158M} by the author's permission}. }
\label{fig:lightcurve+spectra}
\end{figure*}

\textcolor{black}{I) Episode 1 (UPE I): the BH formation caused by hypercritical accretion onto the companion NS in BdHN I, and its subsequent UPE phase originated from vacuum polarization and expanding PEMB pulses with their characteristic Lorentz factor $\Gamma \sim 100$ \citep{1999A&A...350..334R,RSWX2}. This episode pinpoints the first manifestation of the BH (BH-rise) which is the subject of the current paper. The UPE I of GRB 180720B occurs from $~t_{\rm rf}=4.84$~s to $~t_{\rm rf}=6.05$~s. Its measured isotropic energy is $E^{\rm MeV}_{\rm UPE I}=(6.37\pm0.48) \times 10^{52}$~erg, and its spectrum is best fitted by a CPL+BB model (index $\alpha=-1.13$, cutoff energy $E_{\rm c}=2220.569$~keV, and blackbody (BB) temperature $k T = 50.31$~keV in the observer's frame); see Fig.~\ref{fig:lightcurve+spectra}. }

\textcolor{black}{II) Episode 2 ($\nu$NS-rise): the fallback of the SN ejecta onto the $\nu$NS spins it up (\cite{2019ApJ...871...14B} and Becerra et al., submitted). The first evidence of this interplay in GRB 180720B, referred to as the \textit{$\nu$NS-rise}, spans from $~t_{\rm rf}=6,05$~s to $~t_{\rm rf}=9.07$~s. The isotropic energy of this phase $E^{\rm MeV}_{\rm \nu Ns}=(1.13\pm0.04) \times 10^{53}$~erg, and its spectrum is best fitted by a CPL model ($\alpha=-0.98$, and  $E_{\rm c}=737$~keV, in the observer's frame); see Fig.~\ref{fig:lightcurve+spectra}. }

\textcolor{black}{III) Episode 3 (UPE II): it is evidenced by the the first significant observed GeV photon at $~t_{\rm rf}=7.06$~s. This phase also includes the continuation of the UPE phase (UPE II) from $~t_{\rm rf}=9.07$~s to $~t_{\rm rf}=10.89$~s, with an isotropic energy of $E_{\rm UPE II}^{\rm MeV}=(1.6 \pm 0.95) \times 10^{53}$~erg.  A CPL+BB model with model parameters of $\alpha= -1.06^{+0.01}_{-0.01}$, $E_{\rm c}=1502.5^{+88.6}_{-87.5}$~keV and $kT= 39.8^{+1.6}_{-1.6}$~keV is the best fit for the spectrum of this phase; see Fig.~\ref{fig:lightcurve+spectra}.  }

\textcolor{black}{IV) Episode 4 (Cavity): the gravitational collapse of the NS and the consequent BH formation creates a cavity, which becomes further depleted by the expanding PEMB pulses \cite{2019ApJ...883..191R}. The collision and partial reflection of the ultra-relativistic PEMB pulses from the cavity's wall results in radiation with a CPL spectrum that has an energy of $\sim 10^{52}$ erg and a luminosity of $\sim 10^{51}$ erg s$^{-1}$. These values are comparable to the UPE and $\nu$NS-rise energetics \cite{2019ApJ...883..191R}. For GRB 180720B, this emission extends from $t_{\rm rf}=16.94$~s to $~t_{\rm rf}=19.96$~s, with an isotropic energy of $E_{\rm CV}^{\rm MeV}=(4.32 \pm 0.19) \times 10^{52}$~erg, characterized by a CPL spectrum ($\alpha=-1.16$, $E_{\rm c} = 607.96$~keV). Its luminosity and spectrum is given in Fig.~\ref{fig:lightcurve+spectra}.}

\textcolor{black}{V) Episode 5 soft X-ray flare (SXF), and VI) Episode 6 hard X-ray flare (HXF): HXF and SXF emissions result from the interaction of the PEMB pulses with the SN ejecta occurring at $r= 10^{12}$ cm from the BH \cite{2018ApJ...869..151R}.  The HXF of GRB 180720B extends from $t_{\rm rf}= 28.95$~s to $t_{\rm rf}= 34.98$~s, with $L_{\rm HXF,iso}^{\rm MeV}=(7.8 \pm 0.07) \times 10^{51}$~erg~s$^{-1}$. Its spectrum is best fitted by a CPL model with $E_{\rm c}=(5.5_{-0.7}^{+0.8}) \times 10^2$~keV, $\alpha = -1.198 \pm 0.031$. The SXF occurs from $t_{\rm rf}= 55$~s to $t_{\rm rf}= 75$~s, with $L_{\rm SXF,iso}^{\rm X}=1.45\times 10^{50}$~erg s$^{-1}$. Its spectrum is best fitted by a PL+BB model with $\alpha = -1.79 \pm 0.23$, and $k T=0.99 \pm 0.13$~keV; see Fig.~\ref{fig:lightcurve+spectra}.}

\textcolor{black}{The cavity, SXF, and HXF have energetics similar to the UPE phase because they are also created by the interaction of expanding PEMB pulses with SN ejecta; (see, \cite{2021MNRAS.504.5301R}, and references therein).}

%

\textcolor{black}{ One-dimensional \cite{2014ApJ...793L..36F, 2015PhRvL.115w1102F}, two-dimensional \cite{2015ApJ...812..100B}, and three-dimensional \cite{2016ApJ...833..107B, 2019ApJ...871...14B} simulations of BdHN model clearly show that the accretion of the SN ejecta onto the $\nu$NS and NS companion transfers both mass and angular momentum to them. According to numerical simulations of the early evolution phase of BdHN I, the NS companion can reach its critical mass for BH formation before the second peak of fallback accretion onto the $nu$NS (Becerra et al., submitted; see also \cite{2019ApJ...871...14B}). In some cases, this phenomenon allows the $\nu$NS-rise emission to superpose to the UPE. In GRB 180720B, the energy released by the $\nu$NS-rise dominates the UPE phase for about three seconds, resulting in split UPEs I and II. After that, the $\nu$NS-rise emission fades and the UPE becomes visible again. 
}
 
The detailed explanation of Episodes 4 to 6 of GRB 180720B is presented in \cite{2021arXiv210309158M}. \textcolor{black}{This work is devoted to the UPE I and UPE II phases of GRB 180720B. } Following the explanation of the UPE phase in GRB 190114C \cite{PhysRevD.104.063043}, we first present the detailed spectral analysis of the UPE phase of GRB 180720B and then its astrophysical mechanism based on the inner engine of GRBs \cite{2019ApJ...886...82R} \textcolor{black}{and expanding PEMB pulses \citep{2021PhRvD.104f3043M}}.
%

\section{The time-resolved spectral analysis of the UPE phase}\label{sec:heir}

{Due to the high signal-to-noise (S/N) ratio of the Fermi-GBM data acquired during the UPE phase, a refined spectral analysis is performed in the $[4.84$--$6.05]$ time interval in three iterations, and in the $[9.07$--$10.89]$ time interval in five iterations on decreasing time bins, while maintaining reliable statistical significance. The time intervals between iterations are halved.}

{For the final iteration of the UPE I, i.e., the third iteration, the UPE I is divided into four time intervals of $\Delta t_{\rm rf} \approx 0.3$~s: [$4.840$s--$5.142$s],[$5.142$s--$5.445$s],[$5.445$s--$5.747$s], [$5.747$s--$6.050$s].}

For the last iteration of the the UPE II where reliable statistical significance is still fulfilled, i.e., the fifth iteration, the UPE II is divided into $16$ time intervals of $\Delta t_{\rm rf} \approx 0.11$~s: [$9.07$s--$9.19$s],[$9.19$s--$9.30$s],[$9.30$s--$9.41$s], [$9.41$s--$9.53$s], [$9.53$s--$9.64$s], [$9.64$s--$9.75$s], [$9.75$s--$9.87$s], [$9.87$s--$9.98$s], [$9.98$s--$10.10$s], [$10.10$s--$10.21$s], [$10.21$s--$10.32$s], [$10.32$s--$10.44$s], [$10.44$s--$10.55$s], [$10.55$s--$10.66$s], [$10.66$s--$10.78$s] and [$10.78$s--$10.89$s]. 

\begin{figure*}
(a)
\includegraphics[width=0.49\hsize,clip]{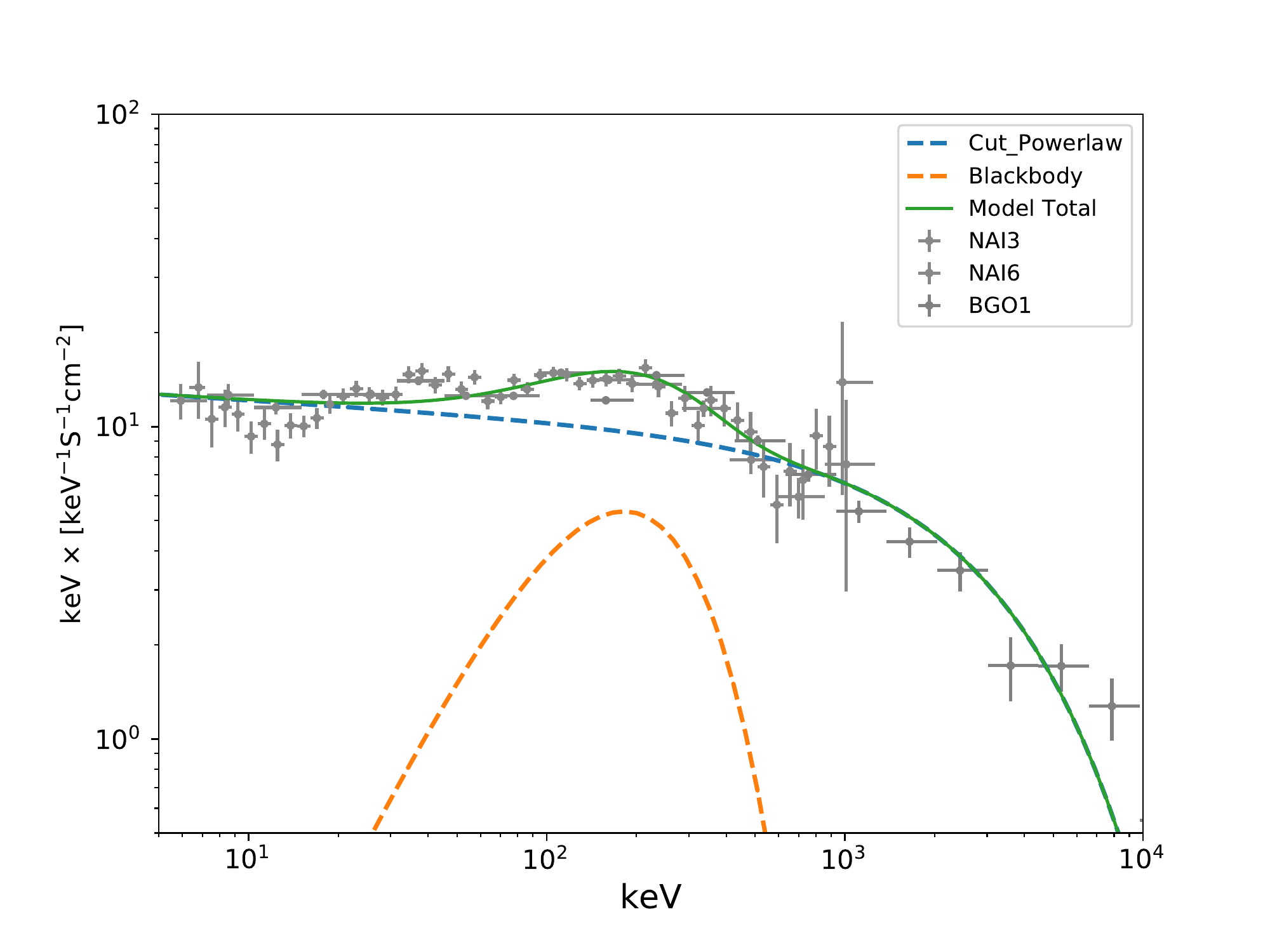}
(b)
\includegraphics[width=0.49\hsize,clip]{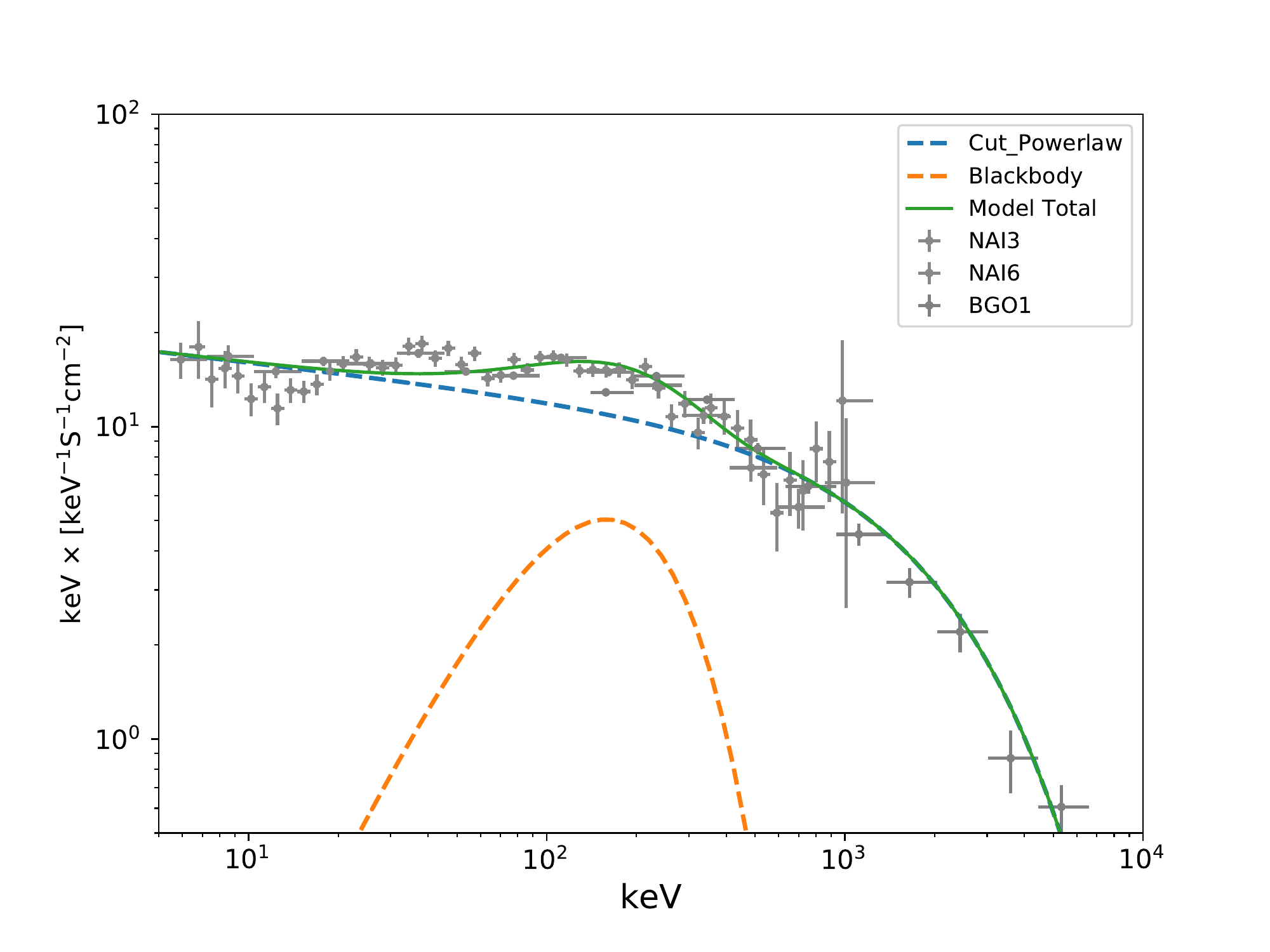}
(c)
\includegraphics[width=0.49\hsize,clip]{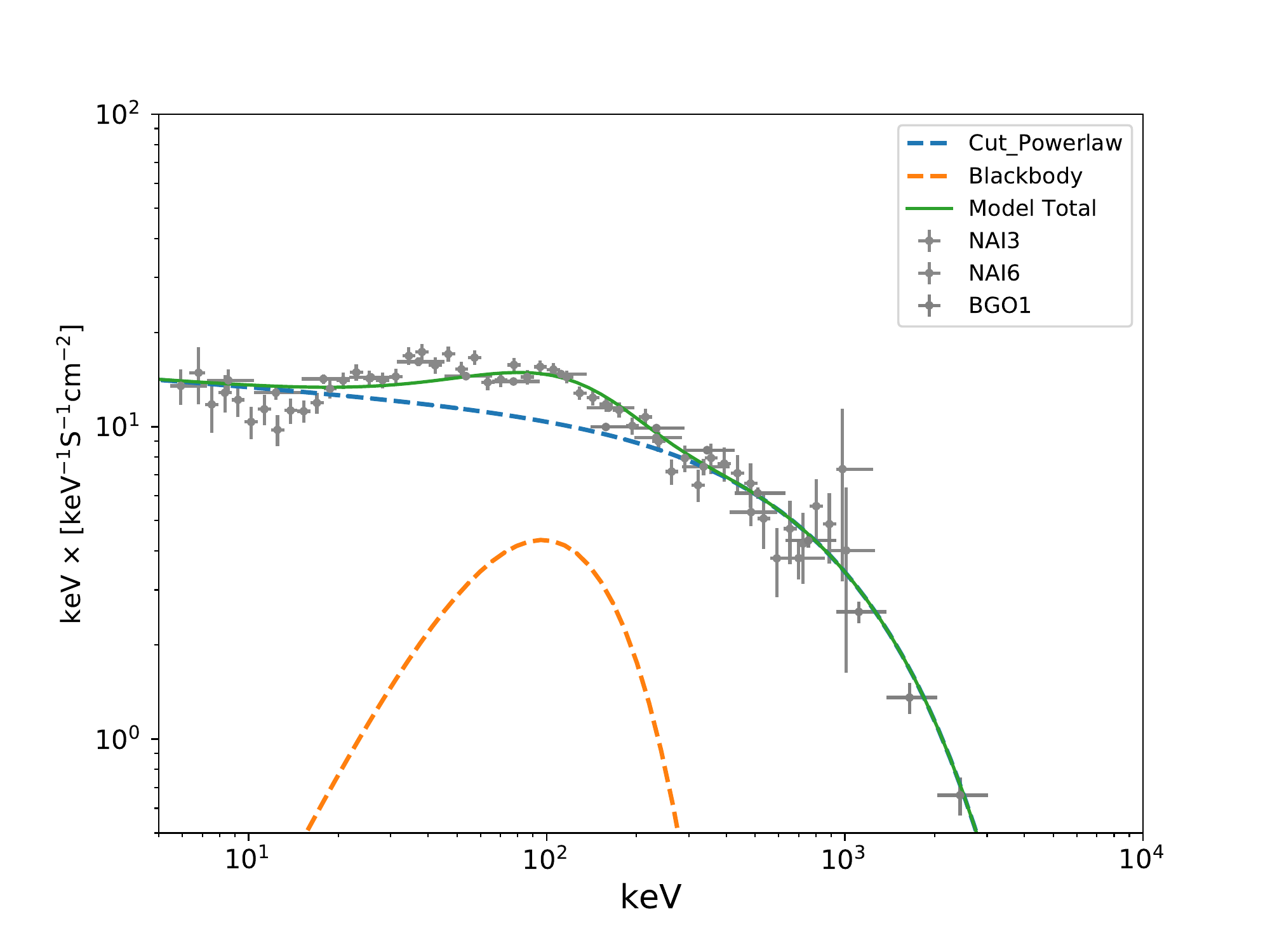}
(d)
\includegraphics[width=0.49\hsize,clip]{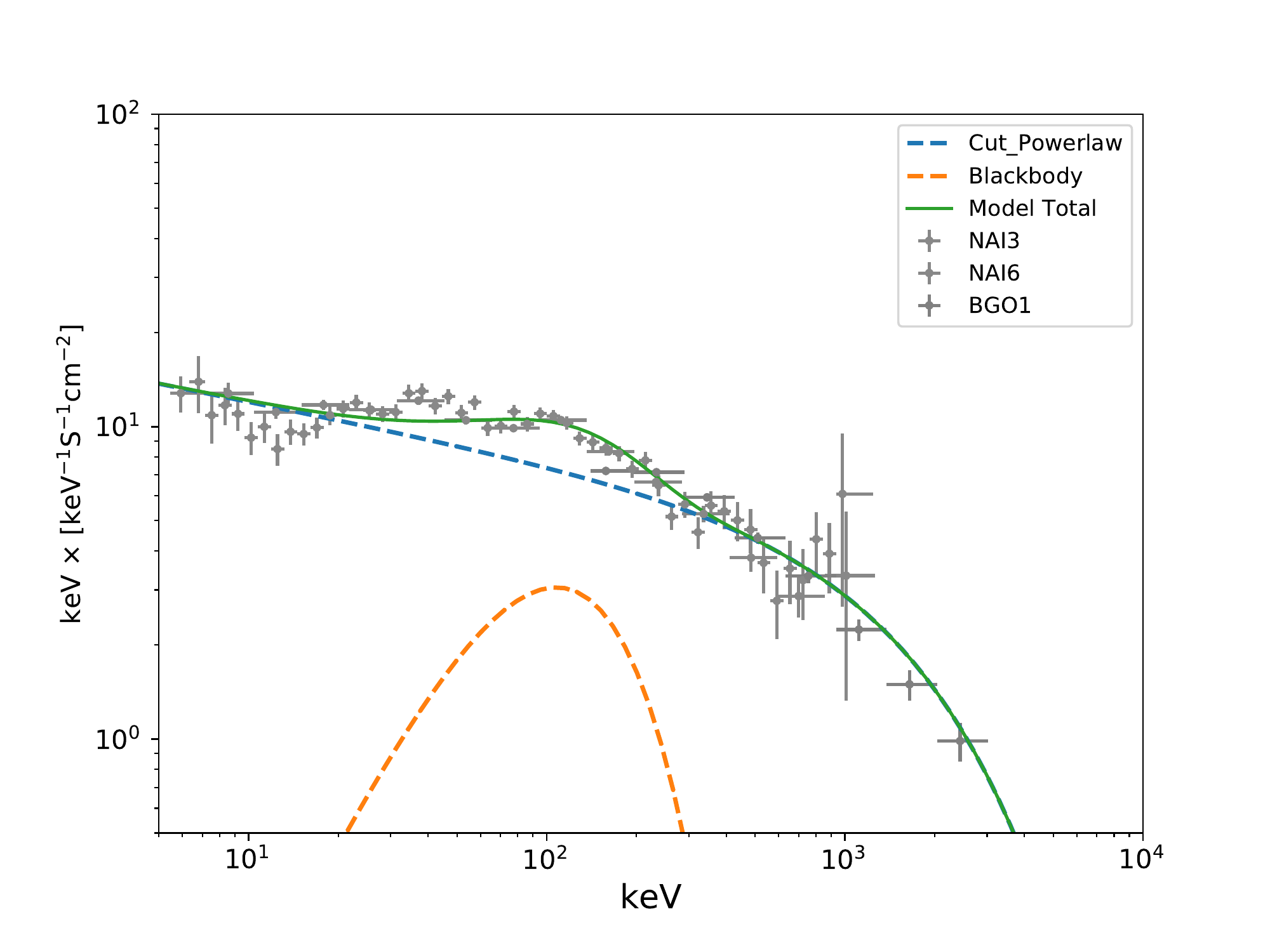}
\caption{{Time-resolved \textcolor{black}{spectral} analysis of the UPE I \textcolor{black}{phase of GRB 180720B}.  \textcolor{black}{This phase, extending from $t_{\rm rf}= 4.84$ to $t_{\rm rf}= 6.05$, is divided into four time intervals with duration of $\Delta t_{\rm rf} \approx 0.3$~s. The spectra of the best fit for the rest-frame time intervals [$4.840$s--$5.142$s], [$5.142$s--$5.445$s], [$5.445$s--$5.747$s], and [$5.747$s--$6.050$s] are shown in panels (a), (b), (c), and (c), respectively. The CPL+BB is confirmed to be the best fit for each time interval}. The best fit parameters are listed in Table~\ref{tab:UPEI}.}}\label{fig:UPEI}
\end{figure*}

\begin{figure*}
\centering
\includegraphics[width=\hsize,clip]{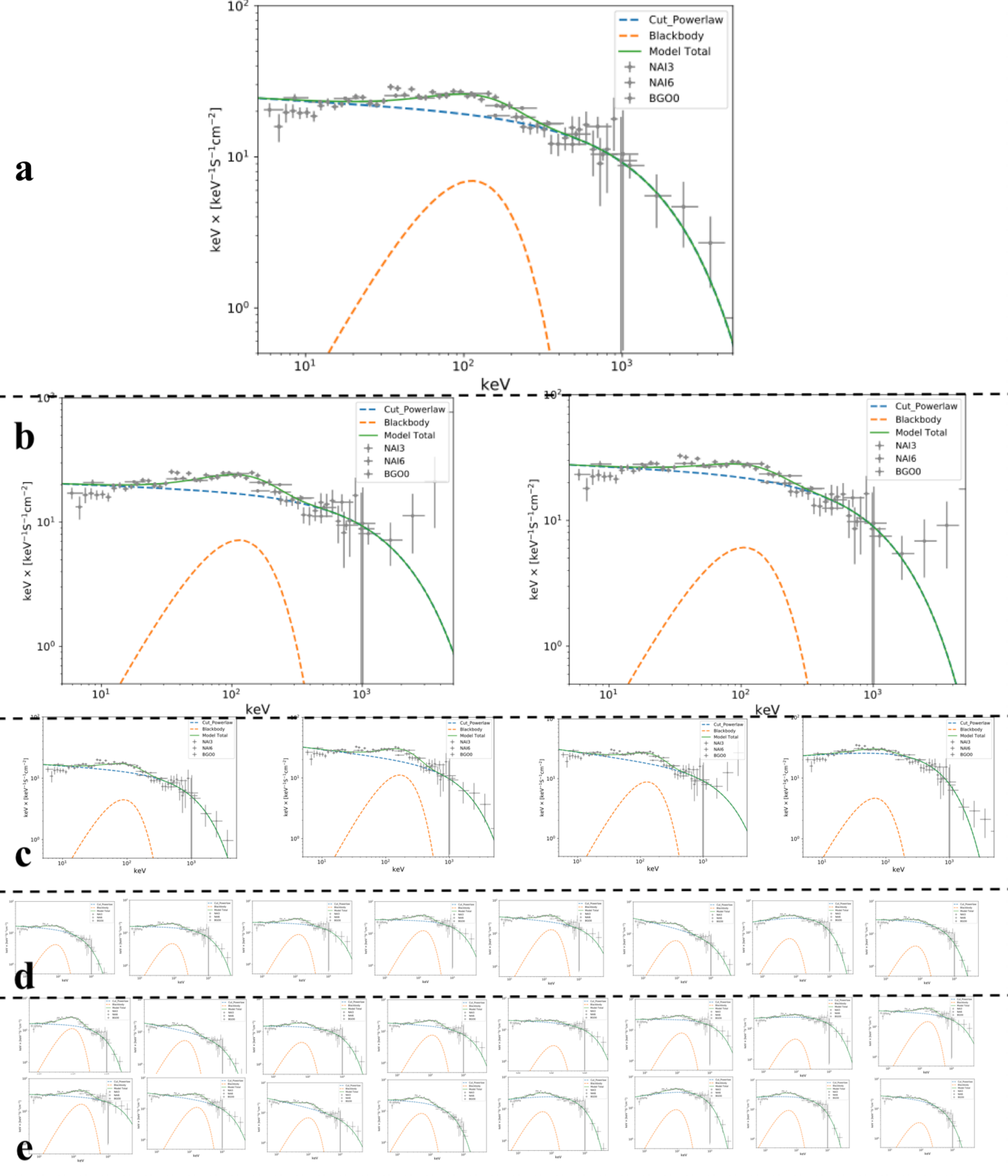}
\caption{Time-resolved spectral analysis of UPE II phase of GRB 180720B. [a]: \textcolor{black}{The first iteration: spectrum of the entire UPE II phase extended from $t_{\rm rf}=9.07$~s to$t_{\rm rf}=10.89$~s.} [b]: \textcolor{black}{The second iteration: spectral analysis carried out} over two equal \textcolor{black}{rest-frame} time intervals of \textcolor{black}{[$9.07$s--$9.98$s] and [$9.98$s--$10.89$s]}.  [c] \textcolor{black}{The third iteration: spectral analysis performed over} four equal \textcolor{black}{rest-frame} time intervals \textcolor{black}{of [$9.07$s--$9.53$s], [$9.53$s--$9.98$s], [$9.98$s--$10.44$s] and [$10.44$s--$10.89$s]. } [d] \textcolor{black}{The fourth iteration: spectral analysis performed} over eighth equal \textcolor{black}{rest-frame} time intervals \textcolor{black}{of [$9.07$s--$9.30$s], [$9.30$s--$9.53$s],[$9.53$s--$9.75$s], [$9.75$s--$9.98$s],[$9.98$s--$10.21$s], [$10.21$s--$10.44$s],[$10.44$s--$10.66$s] and, [$10.66$s--$10.98$s]}. Finally, [e] \textcolor{black}{the fifth iteration: spectral analysis performed} over sixteen equal \textcolor{black}{rest-frame} time intervals \textcolor{black}{of [$9.07$s--$9.19$s],[$9.19$s--$9.30$s],[$9.30$s--$9.41$s], [$9.41$s--$9.53$s], [$9.53$s--$9.64$s], [$9.64$s--$9.75$s], [$9.75$s--$9.87$s], [$9.87$s--$9.98$s], [$9.98$s--$10.10$s], [$10.10$s--$10.21$s], [$10.21$s--$10.32$s], [$10.32$s--$10.44$s], [$10.44$s--$10.55$s], [$10.55$s--$10.66$s], [$10.66$s--$10.78$s] and, [$10.78$s--$10.89$s]. } \textcolor{black}{ The CPL+BB is confirmed to be the best fit for each
time interval.}  The spectral best fit parameters \textcolor{black}{correspond} to each iteration are reported in Table \ref{tab:180720B}.}\label{alltogether}
\end{figure*}

\begin{figure}
\centering
[a]\includegraphics[width=0.9\hsize,clip]{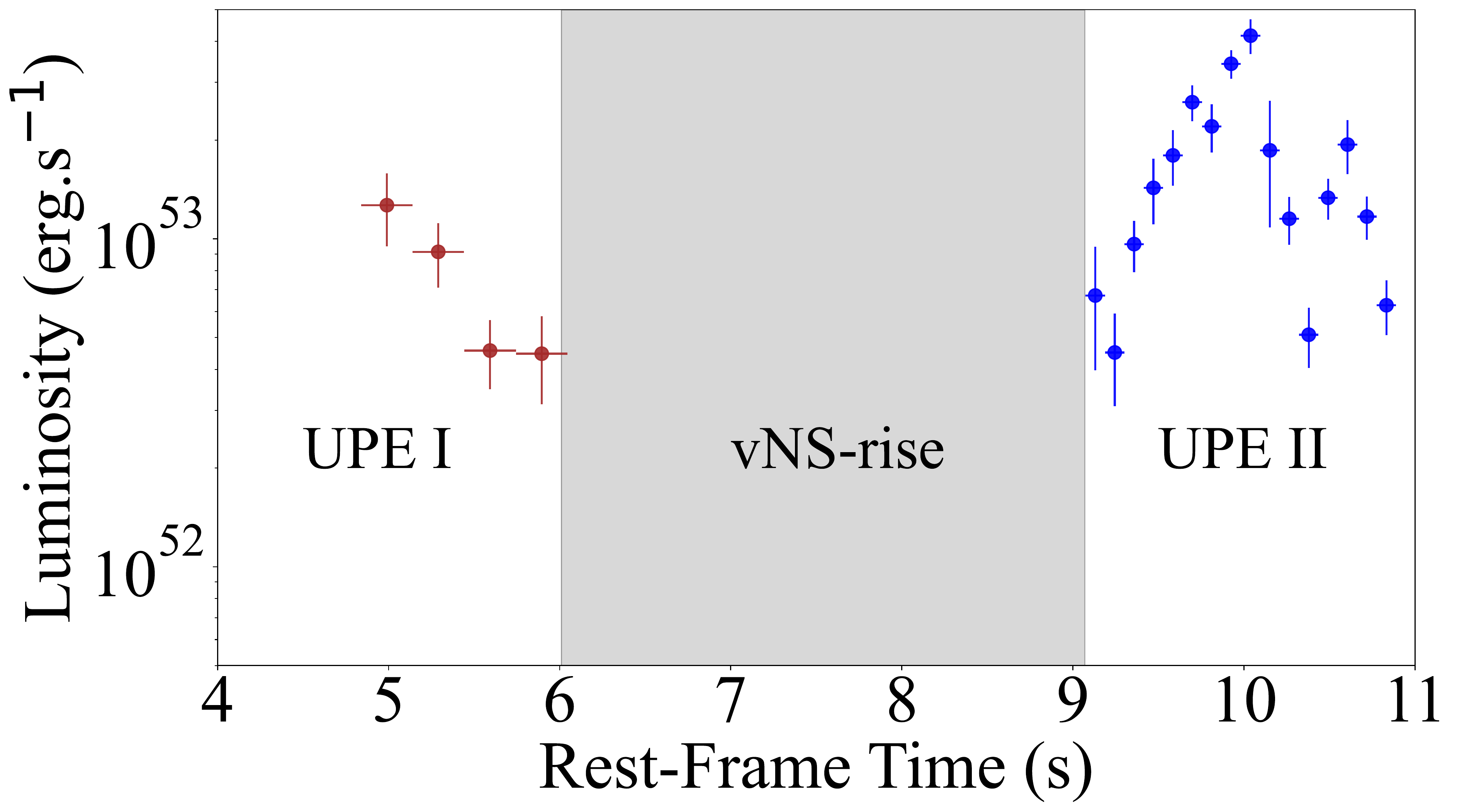}
[b]\includegraphics[width=0.9\hsize,clip]{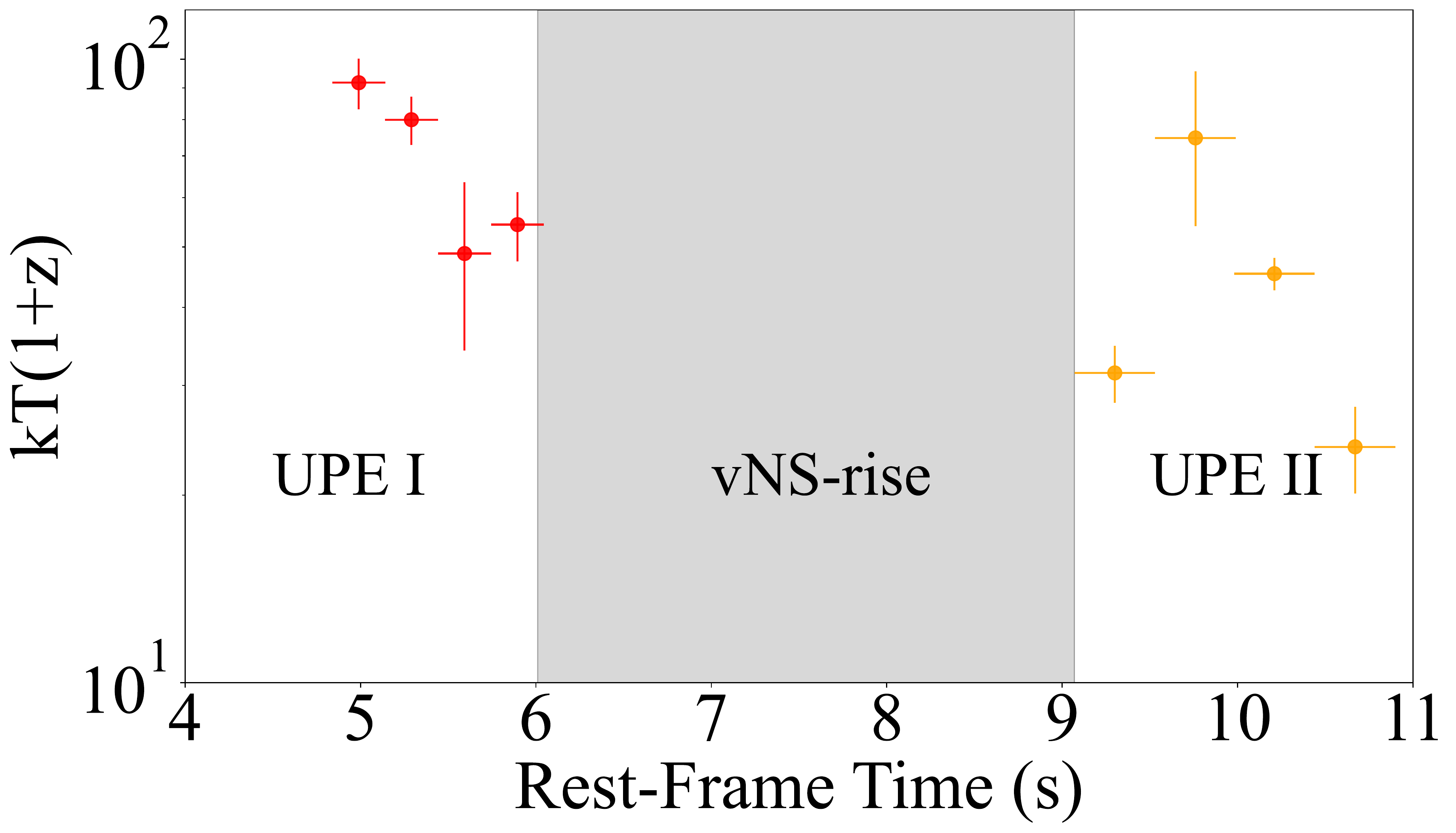}
\caption{[a] The Luminosity during the UPE phase. [b] The rest-frame temperature during the UPE phase. The luminosity and temperature data points are obtained from analyses {with $\Delta t_{\rm rf} =0.3$~s time resolution reported in Table~\ref{tab:UPEI} (UPE I), and } $\Delta t_{\rm rf}=0.11$~s time resolution reported in Table~\ref{tab:180720B} {(UPE II)}.}
\label{fig:lumupe}
\end{figure}

The spectral analysis is performed over each time interval. The presence of a cut-off power-law plus black body (CPL+BB) as the best spectral fit is confirmed in each time interval and for each iterative process. The time intervals both in rest-frame and observer frame, the significance ($S$) for each time interval, the power-law index, cut-off energy, temperature, $\Delta$DIC, BB flux, total flux, the BB to total flux ratio, $F_{\rm BB}/F_{\rm tot}$ and finally the isotropic energy of entire the {UPE phase} and its sub-intervals are shown in {Table~\ref{tab:UPEI} and} Table~\ref{tab:180720B}; see also {Fig.~\ref{fig:UPEI} and} Fig.~\ref{alltogether}. The evolution of the temperature and the luminosity during the UPE phase, as obtained by the time-resolved spectral analysis, are shown in Fig.~\ref{fig:lumupe}.

\begin{table*}
\small\addtolength{\tabcolsep}{-1pt}
\caption{{The time-resolved spectral fit parameters for GRB 180720B (CPL+BB model) during the UPE I phase ($t_{\rm rf}=4.84$ s to $t_{\rm rf}=6.05$ s). This table summarizes the time intervals in the rest and observer frames, their significance ($S$), the power-law index, cut-off energy, temperature, $\Delta$DIC, BB flux, total flux, BB to total flux ratio, $F_{\rm BB}/F_{\rm tot}$, and finally the isotropic energy. The first block (first row) in the table contains the spectral best-fit parameters for UPE I. The second block (second, third, fourth, and fifth rows) contains the time-resolved spectral best-fit parameters for $Delta t_{\rm rf}=0.3$ s.} To select the best model from two different given models, we adopt the deviance information criterion (DIC), defined as DIC=-2log[$p$(data$\mid\hat{\theta}$)]+2$p_{\rm DIC}$, where $\hat{\theta}$ is the posterior mean of the parameters, and $p_{\rm DIC}$ is the effective number of parameters. The preferred model is the model with the lowest DIC score. Here we define $\Delta$DIC=(CPL+BB)-CPL, if $\Delta$DIC is negative, indicating the CPL+BB is better. After comparing the DIC, we find the CPL+BB model is the preferred model over the CPL and other models. } \label{tab:UPEI}

\begin{tabular}{c|c|c|c|c|c|c|c|c|c|c}  
\hline\hline                  
$t_{1}$$\sim$$t_{2}$&$t_{rf,1}$$\sim$$t_{rf,2}$&{$S$}&$\alpha$&$E_{\rm c}$&$kT$&$\Delta$DIC&$F_{\rm BB}$&$F_{\rm tot}$&$F_{\rm ratio}$&$E_{\rm tot}$\\
\hline
(s)&(s)&&&(keV)&(keV)&&(10$^{-6}$)&(10$^{-6}$)&&($10^{52}$~erg)\\
Obs&Rest-frame&&&&&&(erg~cm$^{-2}$~s$^{-1}$)&(erg~cm$^{-2}$~s$^{-1}$)&\\ \hline
8.000$\sim$10.000&4.840$\sim$6.050&142.74&-1.13$^{+0.01}_{-0.01}$&2221$^{+184}_{-183}$&50.3$^{+2.8}_{-2.9}$&-108&1.39$^{+0.53}_{-0.35}$&27.14$^{+2.20}_{-2.04}$&0.05&9.53\\
\hline \hline
8.000$\sim$8.500&4.840$\sim$5.142&82.61&-1.06$^{+0.02}_{-0.02}$&2965$^{+316}_{-313}$&64.4$^{+6.1}_{-6.0}$&-84&2.44$^{+1.43}_{-0.92}$&43.61$^{+4.78}_{-4.80}$&0.06&3.83\\
8.500$\sim$9.000&5.142$\sim$5.445&89.78&-1.11$^{+0.03}_{-0.03}$&1898$^{+266}_{-267}$&56.2$^{+5.0}_{-5.0}$&-51&1.97$^{+1.15}_{-0.75}$&31.47$^{+4.46}_{-4.26}$&0.06&2.76\\
9.000$\sim$9.500&5.445$\sim$5.747&72.53&-1.07$^{+0.06}_{-0.06}$&953$^{+265}_{-267}$&34.2$^{+10.3}_{-13.9}$&-23&0.37$^{+1.53}_{-0.34}$&15.77$^{+7.32}_{-3.93}$&0.02&1.38\\
9.500$\sim$10.000&5.747$\sim$6.050&60.82&-1.19$^{+0.05}_{-0.05}$&1788$^{+571}_{-582}$&38.1$^{+4.9}_{-4.9}$&-32&0.76$^{+0.67}_{-0.39}$&15.42$^{+4.80}_{-4.00}$&0.05&1.35\\
\hline
\end{tabular}
\end{table*}

\begin{table*}
\small\addtolength{\tabcolsep}{-2pt}
\caption{The parameters of the time-resolved spectral best fits of GRB 180720B (CPL+BB model) from the $t_{\rm rf}=9.07$~s to $t_{\rm rf}=10.89$~s. This table reports: the time intervals both in rest-frame and observer frame, the significance ($S$) for each time interval, the power-law index, cut-off energy, temperature, $\Delta$DIC, BB flux, total flux, the BB to total flux ratio, $F_{\rm BB}/F_{\rm tot}$ and finally the isotropic energy. The $\Delta$DICs are reported in column 6. \label{tab:180720B}}             
\centering  

\begin{tabular}{c|c|c|c|c|c|c|c|c|c|c}       
\hline\hline                  
$t_{1}$$\sim$$t_{2}$&$t_{rf,1}$$\sim$$t_{rf,2}$&{$S$}&$\alpha$&$E_{\rm c}$&$kT$&$\Delta$DIC&$F_{\rm BB}$&$F_{\rm tot}$&$F_{\rm ratio}$&$E_{\rm tot}$\\
\hline
(s)&(s)&&&(keV)&(keV)&&(10$^{-6}$)&(10$^{-6}$)&&($10^{52}$~erg)\\
Obs&Rest-frame&&&&&&(erg~cm$^{-2}$~s$^{-1}$)&(erg~cm$^{-2}$~s$^{-1}$)&\\ 
\hline                        

15.00$\sim$18.00&9.07$\sim$10.89&274.60&-1.06$^{+0.01}_{-0.01}$&1502.5$^{+88.6}_{-87.5}$&39.8$^{+1.6}_{-1.6}$&-226.4&1.99$^{+0.43}_{-0.34}$&45.55$^{+3.11}_{-2.70}$&0.04$^{+0.01}_{-0.01}$&16.0$^{+1.1}_{-0.952}$\\
\hline
15.00$\sim$16.50&9.07$\sim$9.98&190.63&-1.04$^{+0.01}_{-0.01}$&1750.5$^{+112.7}_{-111.1}$&40.5$^{+2.0}_{-2.0}$&-176.6&2.08$^{+0.58}_{-0.46}$&48.03$^{+3.28}_{-3.09}$&0.04$^{+0.01}_{-0.01}$&8.46$^{+0.577}_{-0.543}$\\
16.50$\sim$18.00&9.98$\sim$10.89&215.76&-1.05$^{+0.02}_{-0.02}$&1151.3$^{+117.3}_{-119.6}$&37.1$^{+2.8}_{-2.8}$&-78.7&1.63$^{+0.69}_{-0.54}$&41.83$^{+4.61}_{-4.04}$&0.04$^{+0.02}_{-0.01}$&7.37$^{+0.812}_{-0.712}$\\
\hline
15.00$\sim$15.75&9.07$\sim$9.53&105.93&-1.07$^{+0.03}_{-0.03}$&1198.0$^{+211.1}_{-217.8}$&31.4$^{+3.3}_{-3.3}$&-41.5&0.94$^{+0.70}_{-0.42}$&23.84$^{+4.65}_{-3.86}$&0.04$^{+0.03}_{-0.02}$&2.1$^{+0.41}_{-0.34}$\\
15.75$\sim$16.50&9.53$\sim$9.98&163.07&-1.02$^{+0.01}_{-0.01}$&1949.4$^{+126.1}_{-127.9}$&46.2$^{+2.68}_{-2.67}$&-15.4&3.43$^{+1.23}_{-0.84}$&72.68$^{+5.48}_{-5.31}$&0.04$^{+0.01}_{-0.01}$&5.16$^{+0.478}_{-0.423}$\\
16.50$\sim$17.25&9.98$\sim$10.44&155.67&-1.15$^{+0.02}_{-0.02}$&2382.3$^{+217.5}_{-221.3}$&45.3$^{+2.7}_{-2.7}$&-125.6&2.85$^{+1.00}_{-0.76}$&53.96$^{+4.55}_{-4.28}$&0.05$^{+0.02}_{-0.01}$&4.75$^{+0.401}_{-0.377}$\\
17.25$\sim$18.00&10.44$\sim$10.89&159.05&-0.93$^{+0.02}_{-0.02}$&684.7$^{+49.7}_{-49.2}$&23.9$^{+3.8}_{-4.0}$&-30.8&0.63$^{+0.93}_{-0.37}$&35.74$^{+3.28}_{-3.21}$&0.02$^{+0.03}_{-0.01}$&3.15$^{+0.289}_{-0.283}$\\
\hline
15.00$\sim$15.38&9.07$\sim$9.30&69.11&-1.06$^{+0.07}_{-0.08}$&711.2$^{+209.5}_{-215.5}$&28.9$^{+5.7}_{-5.6}$&-30.2&0.78$^{+1.14}_{-0.55}$&14.27$^{+6.80}_{-3.54}$&0.05$^{+0.08}_{-0.04}$&0.628$^{+0.299}_{-0.156}$\\
15.38$\sim$15.75&9.30$\sim$9.53&83.03&-1.01$^{+0.03}_{-0.03}$&1319.4$^{+210.9}_{-208.7}$&31.0$^{+5.2}_{-5.2}$&-28.9&0.83$^{+1.14}_{-0.48}$&32.18$^{+6.45}_{-5.45}$&0.03$^{+0.04}_{-0.02}$&1.42$^{+0.284}_{-0.24}$\\
15.75$\sim$16.12&9.53$\sim$9.75&109.59&-1.02$^{+0.02}_{-0.02}$&1967.9$^{+193.8}_{-194.9}$&43.6$^{+4.0}_{-4.0}$&-72.6&2.63$^{+1.51}_{-0.96}$&62.61$^{+6.83}_{-6.58}$&0.04$^{+0.02}_{-0.02}$&2.76$^{+0.301}_{-0.29}$\\
16.12$\sim$16.50&9.75$\sim$9.98&133.10&-1.01$^{+0.02}_{-0.02}$&1919.4$^{+162.1}_{-168.5}$&47.9$^{+3.5}_{-3.5}$&-107.5&4.31$^{+1.60}_{-1.38}$&82.08$^{+8.46}_{-7.17}$&0.05$^{+0.02}_{-0.02}$&3.61$^{+0.372}_{-0.316}$\\
16.50$\sim$16.88&9.98$\sim$10.21&133.12&-1.09$^{+0.02}_{-0.02}$&2574.3$^{+264.0}_{-267.2}$&55.7$^{+3.8}_{-3.7}$&-117.9&5.16$^{+2.03}_{-1.44}$&83.97$^{+8.79}_{-7.60}$&0.06$^{+0.02}_{-0.02}$&3.7$^{+0.387}_{-0.335}$\\
16.88$\sim$17.25&10.21$\sim$10.44&89.16&-1.24$^{+0.05}_{-0.05}$&1537.9$^{+522.7}_{-558.0}$&31.9$^{+3.4}_{-3.4}$&-27.8&1.38$^{+0.94}_{-0.57}$&24.25$^{+7.37}_{-6.29}$&0.06$^{+0.04}_{-0.03}$&1.07$^{+0.325}_{-0.277}$\\
17.25$\sim$17.62&10.44$\sim$10.66&125.76&-0.86$^{+0.03}_{-0.03}$&696.1$^{+59.2}_{-57.7}$&22.5$^{+3.8}_{-3.7}$&-27.3&0.83$^{+1.39}_{-0.48}$&45.89$^{+5.21}_{-4.69}$&0.02$^{+0.03}_{-0.01}$&2.02$^{+0.23}_{-0.206}$\\
17.62$\sim$18.00&10.66$\sim$10.89&102.97&-1.02$^{+0.04}_{-0.04}$&622.4$^{+77.4}_{-80.6}$&25.7$^{+8.4}_{-9.5}$&-25.5&0.39$^{+1.32}_{-0.34}$&25.51$^{+4.95}_{-3.40}$&0.02$^{+0.05}_{-0.01}$&1.12$^{+0.218}_{-0.15}$\\
\hline
15.00$\sim$15.19&9.07$\sim$9.19&51.57&-1.01$^{+0.14}_{-0.15}$&805.3$^{+449.1}_{-380.0}$&33.0$^{+13.1}_{-18.2}$&-288.8&0.80$^{+5.09}_{-0.77}$&19.23$^{+23.49}_{-7.86}$&0.04$^{+0.27}_{-0.04}$&0.423$^{+0.517}_{-0.173}$\\
15.19$\sim$15.38&9.19$\sim$9.30&42.03&-1.19$^{+0.09}_{-0.09}$&1201.3$^{+667.6}_{-595.4}$&27.5$^{+4.3}_{-4.2}$&-27.1&0.97$^{+1.06}_{-0.55}$&12.89$^{+8.98}_{-4.04}$&0.08$^{+0.1}_{-0.05}$&0.284$^{+0.198}_{-0.0889}$\\
15.38$\sim$15.56&9.30$\sim$9.41&53.84&-1.00$^{+0.04}_{-0.04}$&1158.5$^{+201.4}_{-200.2}$&23.4$^{+8.3}_{-8.3}$&-27.1&0.29$^{+1.66}_{-0.26}$&27.59$^{+7.34}_{-4.93}$&0.01$^{+0.06}_{-0.01}$&0.608$^{+0.162}_{-0.109}$\\
15.56$\sim$15.75&9.41$\sim$9.53&63.61&-1.06$^{+0.05}_{-0.05}$&1839.8$^{+434.0}_{-420.6}$&39.4$^{+7.6}_{-7.0}$&-32.2&1.74$^{+2.60}_{-1.11}$&40.95$^{+11.15}_{-9.24}$&0.04$^{+0.06}_{-0.03}$&0.902$^{+0.246}_{-0.203}$\\
15.75$\sim$15.94&9.53$\sim$9.64&72.54&-1.04$^{+0.04}_{-0.04}$&1896.8$^{+350.9}_{-351.5}$&40.8$^{+4.7}_{-4.7}$&-30.3&2.78$^{+2.13}_{-1.19}$&51.44$^{+12.23}_{-9.91}$&0.05$^{+0.04}_{-0.03}$&1.13$^{+0.269}_{-0.218}$\\
15.94$\sim$16.12&9.64$\sim$9.75&83.99&-0.99$^{+0.03}_{-0.03}$&1950.2$^{+231.8}_{-232.1}$&47.5$^{+7.6}_{-7.6}$&-34.3&2.34$^{+3.12}_{-1.29}$&74.72$^{+11.53}_{-9.35}$&0.03$^{+0.04}_{-0.02}$&1.65$^{+0.254}_{-0.206}$\\
16.12$\sim$16.31&9.75$\sim$9.87&85.09&-0.95$^{+0.04}_{-0.04}$&1379.2$^{+207.4}_{-203.8}$&32.7$^{+5.4}_{-5.3}$&-39.2&1.84$^{+2.29}_{-1.02}$&63.06$^{+12.29}_{-10.56}$&0.03$^{+0.04}_{-0.02}$&1.39$^{+0.271}_{-0.233}$\\
16.31$\sim$16.50&9.87$\sim$9.98&104.94&-1.05$^{+0.02}_{-0.02}$&2304.7$^{+260.1}_{-261.8}$&62.1$^{+2.8}_{-2.8}$&-85.4&6.72$^{+1.63}_{-1.29}$&97.87$^{+12.08}_{-9.75}$&0.07$^{+0.02}_{-0.01}$&2.15$^{+0.266}_{-0.215}$\\
16.50$\sim$16.69&9.98$\sim$10.10&107.18&-1.04$^{+0.03}_{-0.03}$&2737.1$^{+346.9}_{-340.9}$&58.4$^{+5.6}_{-5.6}$&-86.1&6.57$^{+3.89}_{-2.56}$&119.20$^{+16.65}_{-14.38}$&0.06$^{+0.03}_{-0.02}$&2.62$^{+0.367}_{-0.317}$\\
16.69$\sim$16.88&10.10$\sim$10.21&82.58&-1.13$^{+0.13}_{-0.08}$&1910.0$^{+709.1}_{-1074.0}$&58.6$^{+8.6}_{-9.2}$&-86.9&3.67$^{+4.06}_{-3.43}$&53.29$^{+28.29}_{-22.24}$&0.07$^{+0.08}_{-0.07}$&1.17$^{+0.623}_{-0.49}$\\
16.88$\sim$17.06&10.21$\sim$10.32&64.96&-1.24$^{+0.03}_{-0.03}$&2412.4$^{+580.9}_{-576.0}$&34.7$^{+4.0}_{-4.0}$&-28.1&1.52$^{+1.46}_{-0.72}$&32.97$^{+6.96}_{-5.49}$&0.05$^{+0.05}_{-0.02}$&0.726$^{+0.153}_{-0.121}$\\
17.06$\sim$17.25&10.32$\sim$10.44&62.39&-1.06$^{+0.08}_{-0.08}$&480.3$^{+112.6}_{-114.6}$&21.1$^{+8.8}_{-8.9}$&-125.2&0.39$^{+3.01}_{-0.35}$&15.20$^{+8.60}_{-3.47}$&0.03$^{+0.2}_{-0.02}$&0.335$^{+0.189}_{-0.0764}$\\
17.25$\sim$17.44&10.44$\sim$10.55&81.92&-0.89$^{+0.05}_{-0.05}$&720.6$^{+93.9}_{-92.3}$&19.1$^{+3.9}_{-3.8}$&-23.5&0.82$^{+1.62}_{-0.55}$&38.20$^{+8.11}_{-5.42}$&0.02$^{+0.04}_{-0.01}$&0.841$^{+0.179}_{-0.119}$\\
17.44$\sim$17.62&10.55$\sim$10.66&97.68&-0.84$^{+0.05}_{-0.05}$&713.4$^{+96.8}_{-97.0}$&32.3$^{+11.9}_{-10.7}$&-38.1&1.05$^{+5.66}_{-0.87}$&55.49$^{+13.70}_{-10.34}$&0.02$^{+0.1}_{-0.02}$&1.22$^{+0.302}_{-0.228}$\\
17.62$\sim$17.81&10.66$\sim$10.78&82.29&-0.95$^{+0.05}_{-0.05}$&628.7$^{+86.6}_{-86.2}$&19.5$^{+9.9}_{-7.8}$&-66.8&0.33$^{+4.15}_{-0.30}$&33.47$^{+9.11}_{-5.06}$&0.01$^{+0.12}_{-0.01}$&0.737$^{+0.201}_{-0.111}$\\
17.81$\sim$18.00&10.78$\sim$10.89&64.36&-1.08$^{+0.06}_{-0.06}$&565.9$^{+123.9}_{-118.5}$&30.2$^{+7.8}_{-10.3}$&-15.3&0.36$^{+1.63}_{-0.33}$&17.96$^{+6.32}_{-3.42}$&0.02$^{+0.09}_{-0.02}$&0.395$^{+0.139}_{-0.0752}$\\
\hline                                   
\end{tabular}

\end{table*}
The time-resolved spectral analysis over each iteration, reveals a common spectral feature for each time interval characterized by the CPL$+$BB best-fit model with a rest-frame temperature of $kT= 20\sim 60$~keV and the ratio of blackbody flux ($F_{\rm BB}$) to the total flux ($F_{\rm tot}$) of
\begin{equation}\label{eq:ratio} 
0.01\lesssim \frac{F_{\rm BB}}{F_{\rm tot}} \lesssim 0.07.
\end{equation}

{In essence, the UPE II is a continuation of the UPE I, and there is no distinction between the two. The observed discontinuity in the UPE phase is caused by the simultaneous occurrence of the UPE phase and the $\nu$NS—rise in this GRB. The temporal coincidence of these two emissions in a BdHN depends on binary parameters, more relevant the orbital period (Rueda et al.; to be submitted). As a result, we assume in this paper that the UPE phase extends from  $t_{\rm rf}=4.84$ to $t_{\rm rf} =10.89$~s.}

The existence of the BB  components in the spectrum of the UPE phase has been identified as the characteristic signature of $e^+~e^-$ pair creation in presence of baryons (the PEMB pulse) originating from the vacuum polarization process \citep{1999A&AS..138..511R,1999A&A...350..334R, 2000A&A...359..855R,2010PhR...487....1R,PhysRevD.104.063043}. This subject will be addressed in Secs.~\ref{sec:vac} and \ref{sec:12}.

\section{The properties of inner engine}\label{sec:5}

The physics of inner engine was first described in \cite{2019ApJ...886...82R,2020EPJC...80..300R} for GRB 130427A and \cite{2021A&A...649A..75M} for GRB 190114C. The Papapetrou-Wald solution \cite{1966AIHPA...4...83P,1974PhRvD..10.1680W} is used to describe the newborn Kerr BH in the BdHNe I surrounded by a magnetic field and by the low-density plasma in the cavity \citep{2019ApJ...883..191R}. The gravitomagnetic interaction of the newborn Kerr BH with the magnetic field induces a strong electric field in the BH vicinity \cite{2022ApJ...929...56R}. In \cite{PhysRevD.104.063043}, it was shown that the UPE phase of GRB 190114C can originate by the QED process of vacuum polarization in an overcritical field, i.e., $|{\bf E}|>E_c$, where $E_c$ is the critical field for spontaneous $e^+e^-$ pair creation in vacuum and is given by Eq. (\ref{eq:Ec}). Following this framework, we apply in this work the inner engine in the QED overcritical regime to explain the UPE phase of GRB 180720B observed by Fermi-GBM. We give the details of the physical process in Section \ref{sec:vac}.

In this section, we focus on the structure of the electromagnetic field around the BH to investigate the conditions under which the overcritical field regime can develop. The components of the electric and magnetic field (in the Carter's orthonormal tetrad) in the approximation of small polar angles can be written as \citep{2019ApJ...886...82R, 2020EPJC...80..300R}
\begin{eqnarray}
   E_{\hat{r}} &=& -\frac{2 B_0 J\,G}{c^3} \frac{ \left(r^2-\hat{a}^2 \right)}{\left(r^2+\hat{a}^2 \right)^2} \label{eq:ER} \\ 
    E_{\hat{\theta}}&=&0 \\
  B_{\hat{r}}&=&\frac{B_0  \left(-\frac{4\,G\, J^2 r}{M\left(r^2+\hat{a}^2 \right) }+a^2+r^2\right)}{\left(r^2+\hat{a}^2 \right)}\\
  B_{\hat{\theta}}&=& 0.
 \end{eqnarray}
where $\Sigma=r^2+\hat{a}^2\cos^2\theta$, $\Delta=r^2-2 \hat{M} r+\hat{a}^2$, $\hat{M}= G M/c^2$, $\hat{a}=a/c=J/(M\,c)$, being $M$ and $J$ the mass and angular momentum of the Kerr BH. The (outer) event horizon is located at $r_+=(\hat{M}+\sqrt{\hat{M}^2-\hat{a}^2})$. 

We can now introduce the effective charge \cite{2019ApJ...886...82R}
\begin{equation}\label{eq:Qeff}
    Q_{\rm eff}=\frac{G}{c^3}2 B_0 J,
\end{equation}
which when is replaced in the charge of the Kerr-Newman solution, it leads to a radial electric field equal to the one of the Papapetrou-Wald solution given by Eq.~(\ref{eq:ER}) \cite{PhysRevD.104.063043}.
%
Therefore, up to linear order in $\theta$ and in the dimensionless BH spin parameter $\alpha \equiv \hat{a}/(G M/c^2)$, the electric field can be written as
\begin{equation}\label{eq:ER2}
   E_{\hat{r}} = -\frac{2 B_0 J\,G}{c^3} \frac{ \left(r^2-\hat{a}^2 \right)}{\left(r^2+\hat{a}^2 \right)^2}\approx -\frac{1}{2}\alpha B_0\frac{r_+^2}{r^2}.
\end{equation}
%
%
%
 
The specific value of the mass, spin parameter, and magnetic field in the inner engine will be determined as a function of operative astrophysical processes, which are presented in the next sections. We now discuss how the gravitational collapse of the NS in a BdHN I can lead to an engine with an electromagnetic field structure that can be approximately described by the Papapetrou-Wald solution. To the best of our knowledge, there are no numerical simulations dedicated to demonstrating the formation of this specific configuration. Nevertheless, we refer to Sec. $7$ of \cite{2020ApJ...893..148R} which discusses the nature of the magnetic field around the newborn BH in a BdHN I and its support from numerical simulations of gravitational collapse. 

Numerical simulations in \cite{2019ApJ...871...14B} show that the magnetized NS companion of the CO$_{\rm core}$, in the accretion process, gains not only a considerable amount of mass but also angular momentum. Therefore, we conclude that the BH forms from the collapse of a magnetized ($10^{12}$--$10^{13}$ G), fast rotating (millisecond period) NS once it reaches the critical mass. The same simulations show that SN material remains bound around the nascent BH in a torus-like structure. This matter is essential to anchor the magnetic field outside the newborn BH. The matter density in the off-equatorial directions is low, as shown by numerical simulations in \cite{2019ApJ...883..191R}. The above picture naturally leads to the inner engine: a rotating BH surrounded by a magnetic field and very low density ionized matter.

The numerical simulations of the magneto-rotational collapse of NS starting from the seminal simulations of J. Wilson in \cite{1975NYASA.262..123W, 1978pans.proc..644W}. These early works already showed the amplification of the magnetic field in the gravitational collapse. This result is confirmed by simulations of NS binary mergers leading to a BH surrounded by an accretion disk, whose post-merger system show similarities with the inner engine picture and support the present scenario. Some relevant works on this subject are \cite{2006PhRvL..96c1101D, 2006PhRvL..96c1102S, 2006PhRvD..73j4015D, 2007CQGra..24S.207S, 2008PhRvD..77d4001S, 2011ApJ...732L...6R}.

Equation (\ref{eq:ER2}) tells us that if the above processes occur and amplify the magnetic field strength to values $B_0 \gtrsim (2/\alpha) B_c$ near the BH horizon, where $B_c = E_c = 4.41\times 10^{13}$ G, an overcritical electric field will develop and lead to the QED process of vacuum polarization. 

\section{Mass and spin of BH}\label{sec:massupe}

The energy condition is obtained from the mass-energy formula of the Kerr BH \cite{1970PhRvL..25.1596C,1971PhRvD...4.3552C,1971PhRvL..26.1344H}
%
\begin{equation}
\label{aone}
M^2 = \frac{c^2 J^2}{4 G^2 M^2_{\rm irr}}+M_{\rm irr}^2.
\end{equation}

The extractable energy of a Kerr BH $E_{\rm ext}$ {is given by the subtracting the irreducible mass, $M_{irr}$, from  the total mass of the BH, $M$}:
\begin{equation}
\label{Eextr}
E_{\rm ext}=(M-M_{\rm irr}) c^2=\left(1-\sqrt{\frac{1+\sqrt{1-\alpha^2}}{2}}\right)M c^2.
\end{equation}
which we use  to obtain $M$ as a function of $\alpha$, $M(\alpha)$, by requesting the condition that observed UPE emission originates from BH extractable energy, i.e.
\begin{equation}
\label{latextr}
E_{\rm UPE} = E_{\rm ext}.
\end{equation}

The goal is to show that the Kerr BH extractable energy can explain the energetics of the UPE phase and, in turn, it leads to estimate the mass and spin of BH. Equation (\ref{Eextr}) has two parameters, $M$ and $\alpha$, hence we must supply another equation to determine them.

\begin{figure}
\centering
\includegraphics[width=\hsize,clip]{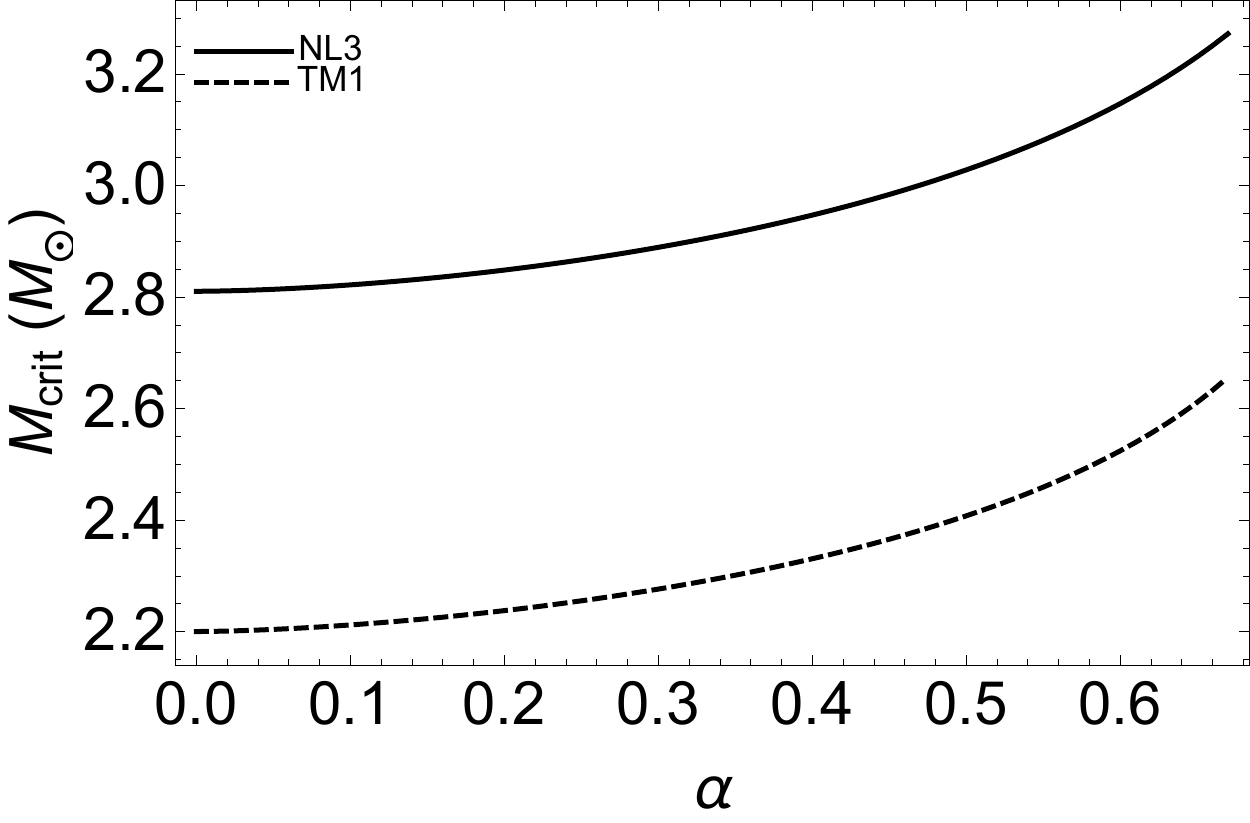}
\caption{NS critical mass as a function of the spin parameter $\alpha$ for the NL3 and TM1 EOS. We recall that the maximum spin parameter of a uniformly rotating NS is $\alpha_{\rm max}\approx0.71$, independently of the NS EOS; see e.g. \citet{2015PhRvD..92b3007C}.}
\label{fig:Mcrit}
\end{figure}

In BdHNe I the BH originates from the hypercritical accretion of SN ejecta onto the NS, i.e., the BH forms when the NS reaches its critical mass. Therefore, the mass of the BH must satisfy the constraint
\begin{equation}
\label{meq}
M\geq M_{\rm crit}(\alpha),
\end{equation}
where $M_{\rm crit}(\alpha)$ is the critical mass of a rotating NS with Kerr spin parameter $\alpha$. The NS critical mass value depends on the nuclear equation of state (EOS). In \citet{2015PhRvD..92b3007C}, it was shown that, for instance, for the NL3, GM1 and TM1 EOS, the critical mass for rigidly rotating NS is fitted with a maximum error of $0.45\%$ by the expression
\begin{equation}\label{eq:Mcrit}
M_{\rm crit}(j)=M_{\rm crit}^{J=0}(1 + k j^p),
\end{equation}
where $k$, $p$, and $M_{\rm crit}^{J=0}$ are parameters that depend upon the nuclear EOS, being the latter the critical mass in the non-rotating case, and $j \equiv cJ/(G M_\odot)^2$. With the relation between $j$ and $\alpha$, i.e., $j = \alpha (M/M_\odot)^2$, Eq. (\ref{eq:Mcrit}) becomes an implicit non-linear algebraic equation for the NS critical mass as a function of $\alpha$. For instance, we show in Fig.~\ref{fig:Mcrit} the numerical solution of Eq. (\ref{eq:Mcrit}) for the NL3 ($k=0.006$, $p=1.68$) and TM1 ($k=0.017$, $p=1.61$) EOS. We limit the value of the spin parameter to $\alpha_{\rm max}\approx 0.7$, which has been found to be the maximum value attainable by rigidly rotating NS independent on the nuclear EOS (see \cite{2015PhRvD..92b3007C} for details). 

We now proceed to estimate the mass and spin parameter of the BH at the beginning of the UPE phase, namely at $t_{\rm rf}=4.84$~s. For this task, we solve Eq. (\ref{latextr}) using $E_{\rm UPE}=E_{\rm UPEI}+E_{\rm UPEI}=2.24 \times 10^{53}$ erg, together with the inequality (\ref{meq}). We use the minimum possible value in the latter so to set a lower limit to the BH mass, and correspondingly an upper limit to the spin parameter. For the NS critical mass, we use  Eq.~(\ref{eq:Mcrit}) for the TM1 EOS. We obtain the lower limit to the BH mass, $M=2.40 M_\odot$, and the upper limit to the spin, $\alpha = 0.60$. The corresponding irreducible mass of the BH which is assumed to be constant during the radiation process is $M_{\rm irr}=2.28~M_\odot$. 


Since the MeV emission during the UPE phase is powered by the extractable rotational energy of the Kerr BH [see Eq. (\ref{latextr})], the time derivative of Eq.~(\ref{Eextr}) gives the luminosity
\begin{equation}
\label{sdown1}
L_{\rm UPE}=-\frac{dE_{\rm ext}}{dt}=-\frac{dM}{dt}.
\end{equation}

\begin{figure}
    \centering
\includegraphics[width=8 cm]{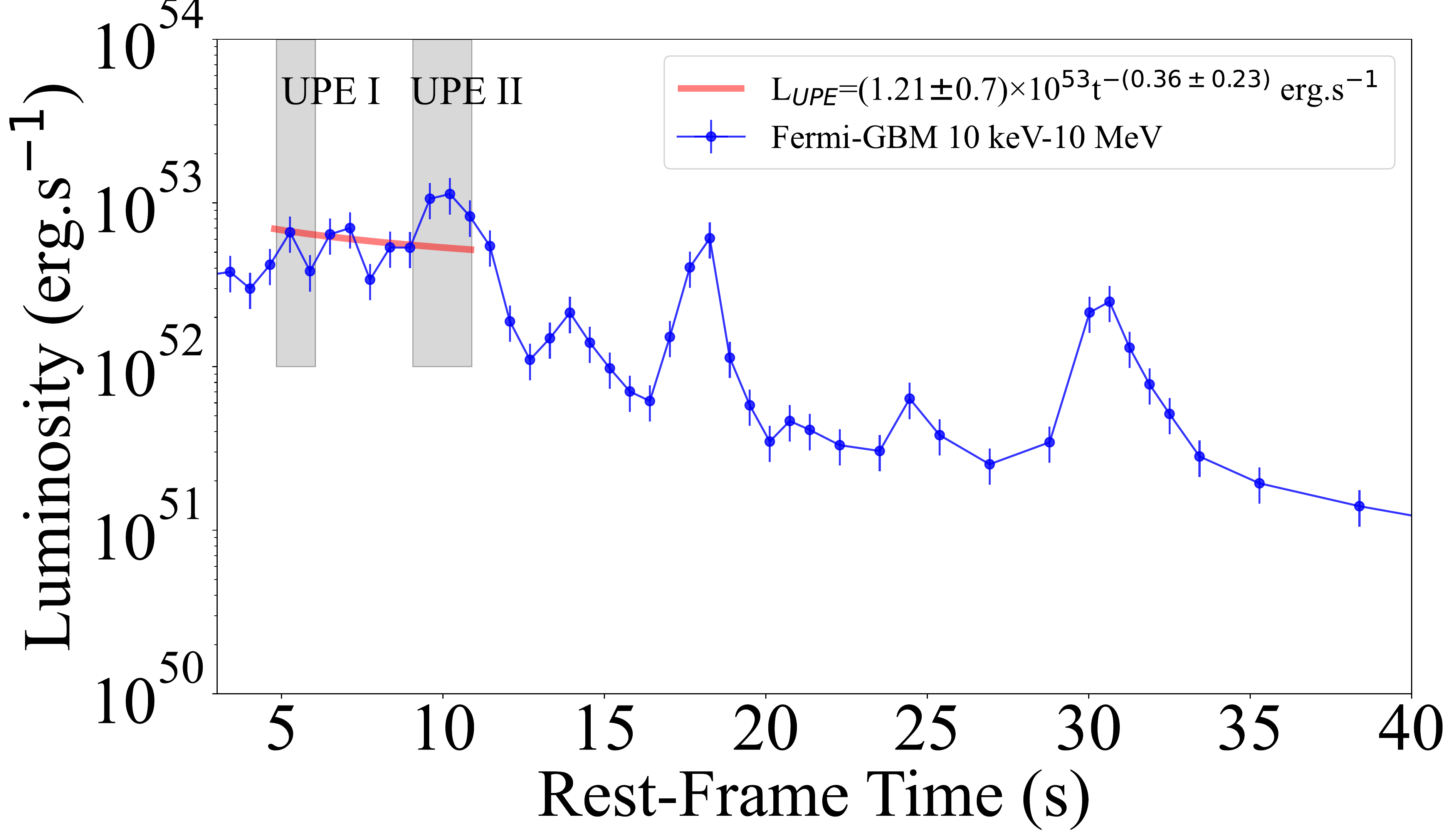}
     \caption{Luminosity of GRB 180720B in the cosmological rest-frame of the source. Blue circles: obtained from \textit{Fermi}-GBM in the $10$~keV--$10$~MeV energy band.  The light grey parts represent the UPE phase; UPE I from $~t_{\rm rf}=4.84$~s to $~t_{\rm rf}=6.05$~s, and UPE II from $~t_{\rm rf}=9.07$~s to $~t_{\rm rf}=10.89$~s.  The rest-frame luminosity light-curve of GRB 180720B during the UPE phase is fitted by a power-law with slope of $\alpha_{\rm UPE}=0.36\pm 0.23$ and, amplitude of $(1.21\pm 0.70)\times 10^{53}$~erg~s$^{-1}$. }\label{fig:GeV+MeV}%
\end{figure}

\begin{figure}
\centering
[a]\includegraphics[width=7.7cm]{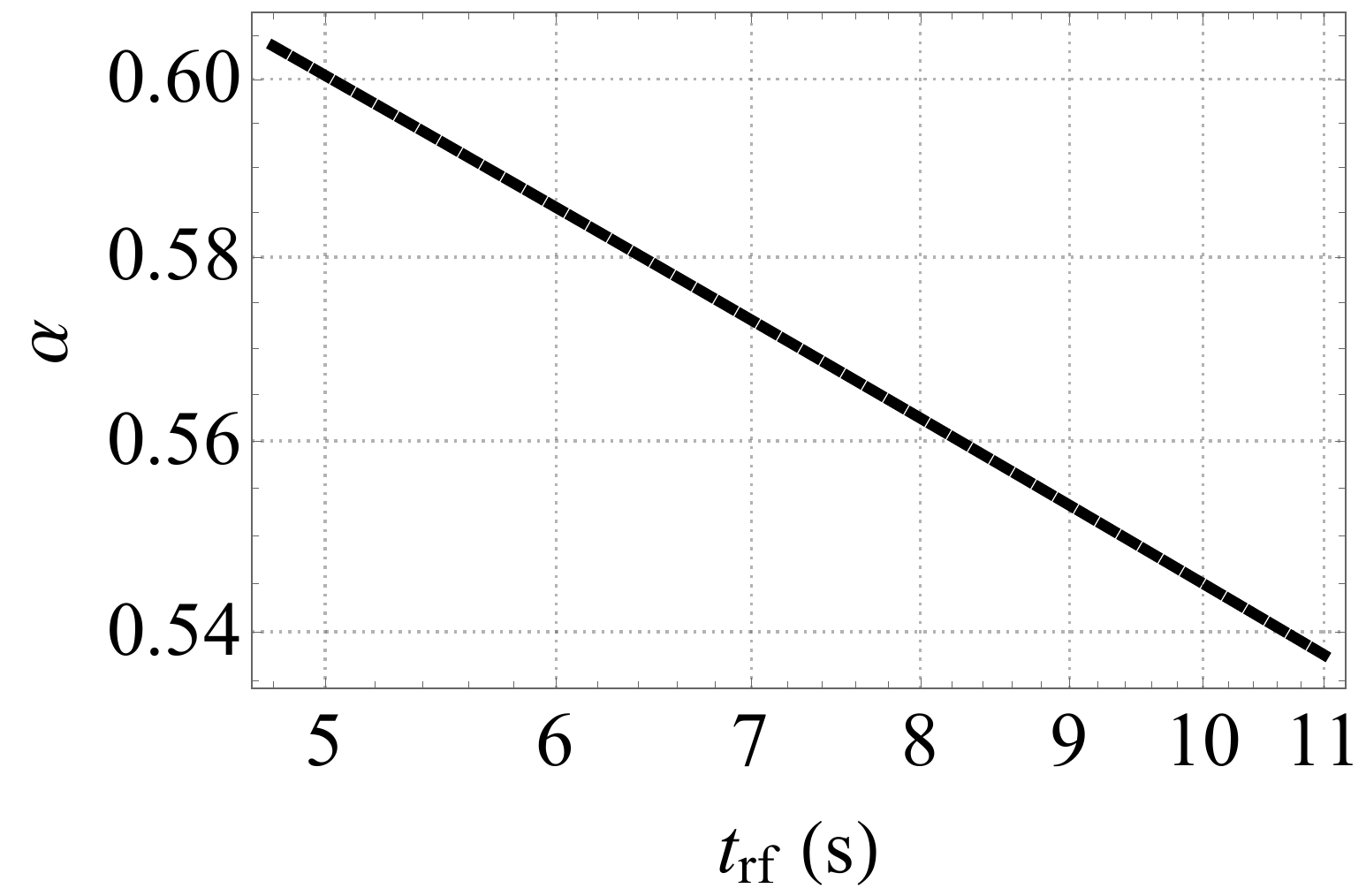}
[b]\includegraphics[width=7.8cm]{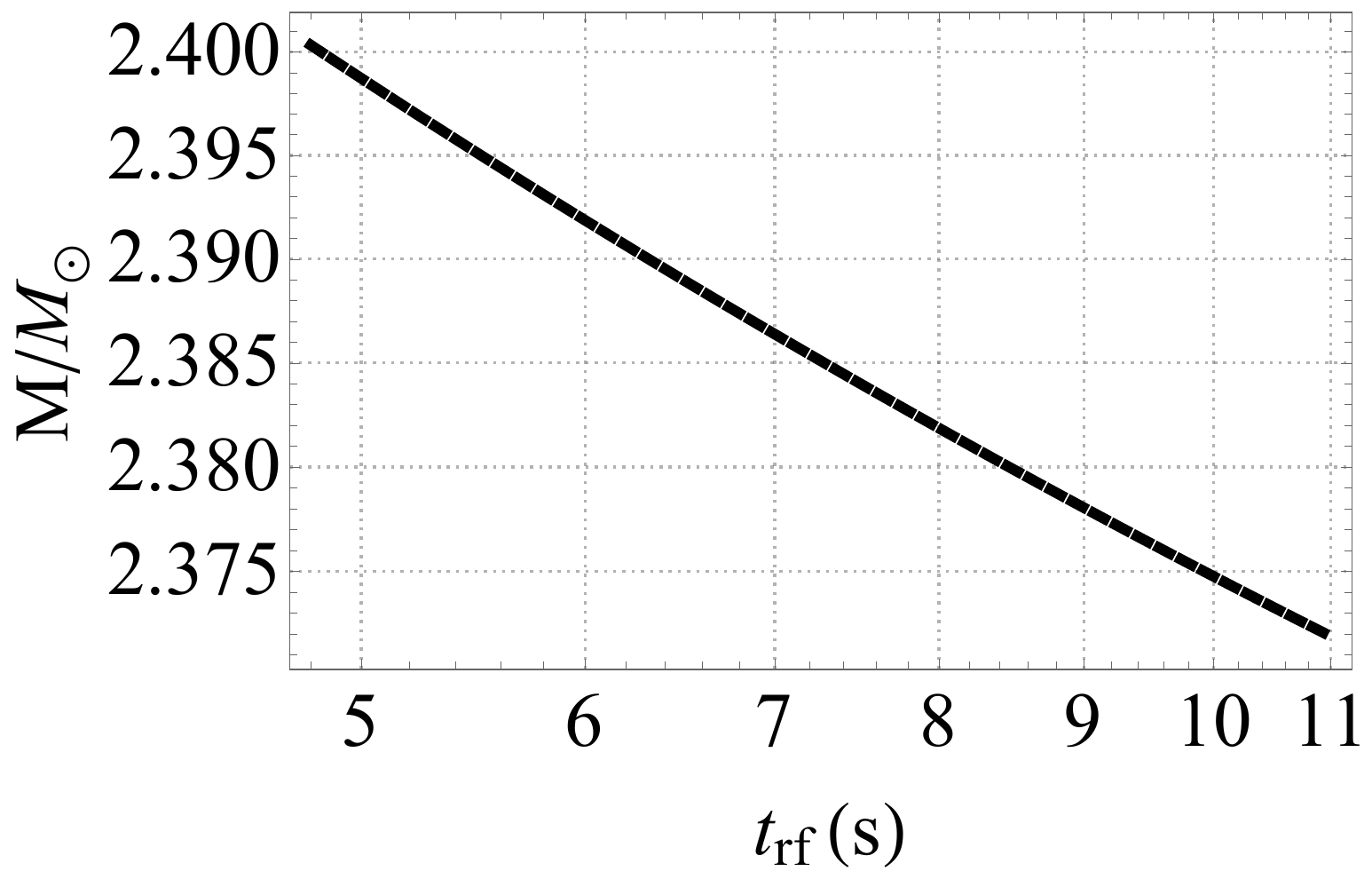}
\caption{ 
The evolution of the mass and spin of BH, as a function of rest-frame time (t$_{\rm rf}$) for GRB 180720B during the UPE phase ({$t_{\rm rf}=4.84$--$10.89$~s}). The MeV energetic is paid by the exctractable energy of the Kerr BH (Eq.~\ref{sdown1}), leading to the reduction of the mass and spin during the UPE phse. }
\label{massspinupe}
\end{figure}

The rest-frame $10$~keV--$10$~MeV luminosity light-curve of GRB 180720B during UPE phase is fitted by a power-law with slope of $\alpha_{\rm UPE}=0.36\pm 0.23$ and, amplitude of $(1.21\pm 0.70)\times 10^{53}$~erg~s$^{-1}$. From this luminosity (see Fig.~\ref{fig:GeV+MeV}), and using as initial BH mass and spin at $t_{\rm rf}=4.84$~s the values estimated above, Eq.~(\ref{sdown1}) can be integrated to obtain the time evolution of the BH mass and spin during the UPE phase which is shown in Fig.~\ref{massspinupe}.

\section{Vacuum polarization and its role in the formation of UPE phase} \label{sec:vac}

The UPE phase has been explained by introducing the concept of the the dyadosphere around the Reissner-Nordström BH \citep{1998astro.ph.11232R} and dyadotorous around the Kerr-Newman geometry \cite{2009PhRvD..79l4002C}.  The dyadoregion is the region around the BHs, characterised by the overcritical electric field $|{\bf E}|> E_c$, filled by the highly dense $e^+~e^-$ pairs produced by vacuum polarization process (see \cite{2010PhR...487....1R} for more details). 

Following Ref. \cite{PhysRevD.104.063043}, we use the dyadoregion framework in the Kerr-Newman geometry taking take advantage of the effective charge defined by Eq. (\ref{eq:Qeff}). For the calculation of the transparency properties of the $e^+e^-$ pair plasma formed from the vacuum polarization process (analyzed in Section \ref{sec:12}), we need the initial energy of the pairs, $E_{e^+e^-}$, and the radius of the dyadoregion, $r_d$. 

The dyadoregion extends from the BH horizon to the distance $r_d$ at which the electric field has the critical value $E_c$. Applying this condition to the Kerr-Newman electric field one obtains \cite{2009PhRvD..79l4002C}
\begin{equation}
\label{dyadosurf}
\left(\frac{r_d}{\hat{M}}\right)^2=\frac12\frac{\lambda}{\mu\epsilon}
-\alpha^2+\left(\frac14\frac{\lambda^2}{\mu^2\epsilon^2}
-2\frac{\lambda}{\mu\epsilon}\alpha^2\right)^{1/2}\,
\end{equation}
with $\epsilon=  E_c M_\odot G^{3/2}/c^4\approx 1.873 \times10^{-6}$, and 
\begin{equation}
\label{eq:lamda}
\lambda = \frac{Q_{\rm eff}}{\sqrt{G} M} = \frac{2 B_0 J G/c^3}{\sqrt{G} M},    
\end{equation}
is the effective charge-to-mass ratio. Therefore, the width of the dyadoregion is
\begin{equation}
\label{eq:width}
 \Delta_{\rm d}(t)= r_d(t)-r_+(t).
\end{equation}

The energy of $e^+~e^-$ pairs generated (at a given time) by the inner engine is estimated as the electromagnetic energy stored in the dyadoregion (see \cite{2009PhRvD..79l4002C} for details), i.e., $E_{e^+e^-} = E_{(r_+,r_d)}$, where
\begin{eqnarray}
\label{Eemxi}
E_{(r_+,r_d)}
&=&\frac{(2 B_0 J G/c^3)^2}{4r_+}\left(1-\frac{r_+}{r_{\rm d}}\right)+\frac{(2 B_0 J G/c^3)^2}{4\hat{a}}\nonumber\\
&\times&\left[\left(1+\frac{\hat{a}^2}{r_+^2}\right)\arctan\left(\frac{\hat{a}}{r_+}\right)\right.\nonumber\\
&-&\left.\left(1+\frac{\hat{a}^2}{r_d^2}\right)\arctan\left(\frac{\hat{a}}{r_d}\right)\right].
\end{eqnarray}

%
%

\section{General formulation of transparency condition of the UPE phase}\label{sec:12}

We follow the treatment of the transparency introduced in \cite{PhysRevD.104.063043}. The existence of the overcritical electric field around the BH leads to the following sequence of events:

1) The formation of an optically thick dyadoregion around BH dominated by the high density and pressure of the neutral $e^+e^-\gamma$ plasma \citep{2010PhR...487....1R}, formed in a timescale $\sim \hbar/(m_e c^2) \approx 10^{-21}$ s, with total energy $E_{e^+~e^-}^{\mathrm{tot}}=E_{\rm iso}$. This plasma is endowed with a baryonic mass $M_B$, with baryon load parameter ${\cal B}=M_B c^2/E_{\rm iso}$. This optically thick pair electromagnetic-baryon pulse is known as the PEMB pulse first introduced by \cite{1999A&A...350..334R}.

2) The self-acceleration and expansion of the PEMB pulses due to their high internal pressure achieved by pair-plasma thermalization in a very short timescale ($\sim 10^{-13}$~s). They reach ultra-relativistic velocities of up to $\Gamma \sim 100$ in the case of long GRBs \cite{1999A&A...350..334R,2007PhRvL..99l5003A,2009PhRvD..79d3008A}).

3) Emission of thermal radiation. When the PEMB pulses expand with ultra-relativistic velocities, the $e^+~e^-~\gamma$ plasma becomes optically thin \cite{1999A&A...350..334R,2000A&A...359..855R}. The condition of transparency is
\begin{eqnarray}
\label{eq:transp1}
\tau &=&  \sigma_T (n_{e^+e^-}+\bar{Z} n_B)\Delta_{d} \approx \sigma_T (\bar{Z} n_B) \Delta_{d},\nonumber \\
& = & \sigma_T \frac{\bar{Z} M_B }{m_N 4 \pi R^2  \Delta_{d}} \Delta_{d}  = 1,
\end{eqnarray}
where $\Delta_{d}$ is the thickness of the PEMB pulses, $\sigma_T$ is the Thomson cross-section, $\bar{Z}$ is the average atomic number of baryons ($\bar{Z}= 1$ for Hydrogen atom and $\bar{Z}= 1/2$ for general baryonic matter), $m_N$ is nucleon mass  and $M_B$  is the baryon mass. For the values of ${\cal B}$ considered in the present work, i.e., ${\cal B}=10^{-3}$--$10^{-2}$, we can safely assume $n_{e^+e^-} \ll n_B$. Therefore, from Eq. (\ref{eq:transp1}) the lower bound of transparency radius is 
\begin{equation}
R^{\rm tr} = \left(\frac{\sigma_T}{8 \pi} \frac{M_B }{m_N}\right)^{1/2}=\left(\frac{\sigma_T}{8\pi}\frac{{\cal B} E_{\rm iso}}{m_Nc^2}\right)^{1/2}.
\label{eq:transp2}
\end{equation}

This emission at transparency, previously known as P-GRB, is characterised by a thermal component observed in the spectral analysis of prompt emission of GRBs. The energy of this blackbody component that signs the occurrence of the UPE phase is
\begin{eqnarray}
\label{eq:energy3}
E^{obs}_{\rm P-GRB} = a T^4_{obs} \Gamma^4(1-v/c)^3  4\pi R_{\rm tr}^2 \Delta_{d},
\end{eqnarray}
where $a=4\sigma/c$, being $\sigma$ the Stefan-Boltzmann constant.

\textcolor{black}{The most efficient process to create the $e^+e^-$ plasma around BH is the vacuum polarization, which proceeds on a quantum timescale of the order of the Compton time, $\hbar/(m_e c^2) \approx 10^{-21}$~s \citep{RSWX2}. The electric field screening time is given by the time it takes to charged particles to induce a field that opposites to the original field. This timescale, of the order of $r_+/c \approx 10^{-5}$~s  ($r_+$ is the BH horizon radius), is $16$ orders of magnitudes larger than quantum time scale. } This guarantees that the formation of the $e^+e^-$ pair plasma and its self-expansion by internal pressure starts before any screening process of the electric field could be at work. The dynamics of the expanding plasma from the vicinity of the BH up to the transparency point depends upon the plasma energy, $E_{e^+e^-}^{\mathrm{tot}}$, and the baryon load parameter, ${\cal B}$ \cite{1999A&A...350..334R,2000A&A...359..855R}.

\textcolor{black}{As discussed in Sec. XI in \citep{2021PhRvD.104f3043M} and in Secs. \ref{sec:magnetic} and \ref{sec:conc} of this paper, the BH extractable energy powers the energy for the creation of the $e^+e^-$ plasma around the BH, which is then used in the kinetic energy of expansion of the PEMB pulse and in the radiation released at transparency. Therefore, in each of these processes, the Kerr BH loses a fraction of its mass-energy and angular momentum. This implies that the BH mass and angular momentum, at the time $t_0+\Delta t$, are $M = M_0- \Delta M$ and  $J = J_0 - \Delta J$, where $\Delta M$ and $\Delta J$ are the BH mass-energy and angular momentum extracted by the PEMB pulse expansion and emission process. We estimate that each process extracts $\Delta M/M \sim \Delta J/J\sim 10^{-9}$. Since the induced electric field depends linearly on $J$, see Eq. (\ref{eq:ER}), the new value of the induced electric field, $E$, is lower than the previous value, $E_0$, fulfilling $E = E_0 (1-\Delta J/J)$. As a result, the system begins a new process in presence of the same magnetic field $B_0$, which is kept constant, and a new, lower effective charge $Q_{\rm eff}= Q_{{\rm eff},0}-\Delta Q_{\rm eff}$, where $\Delta Q_{\rm eff} =2 B_0 \Delta J$. Therefore, we assume that the spacetime evolves from one stationary axially symmetric metric to the next, and at each step the electromagnetic field structure of the inner engine is given by the Papapetrou-Wald solution, and the latter can be approximated by the Kerr-Newman metric of charge $Q_{\rm eff}$. Once the plasma is formed, it self-accelerates, expanding to the point of transparency. We recall that the dynamics of the plasma depends only on the initial conditions of energy and baryon load, which in turn depend only on $M$, $J$ and $Q_{\rm eff}$. This means that the plasma dynamics at times $t>t_0$, being $t_0$ the time of its formation and beginning of the expansion, depends only on the values at $t_0$ of $M$, $J$ and $B_0$. Thus, the QED process of $e^+e^-$ formation and its dynamics leading to transparency of the PEMB pulses efficiently extracts the BH energy without being affected by any screening process of the electric field. Therefore, the decrease of the electric field with time is driven by the BH energy extraction which lowers the BH angular momentum, not because of an electric field screening. This ensures that the above process can repeat over time until the electric field reaches the critical value. For GRB 180720B, this occurs at $t_{\rm rf}=10.89$~s. After this time, the vacuum polarization process does not occur any longer.}

The corresponding value of the Lorentz factor at the instant of transparency, $\Gamma$, and the baryon load parameter can be inferred from UPE observables as follows. The calculation involves the following quantities: (a) the isotropic energy of PEMB pulses, $E_{\rm iso}$; (b) the ratio of the blackbody energy of the P-GRB to the isotropic energy, $E^{\rm obs}_{\rm P-GRB}/E_{\rm iso}$; (c) the observed value of the blackbody temperature of the P-GRB, $T_{\rm obs}$; (d) the width of the dyadoregion at decoupling, $\Delta_{\rm d}$, obtained from Eq.~(\ref{eq:width}). The properties of the plasma at transparency are obtained from the solution of the following equations simultaneously. 

The first equation is obtained by substituting Eq.~(\ref{eq:transp2}) into Eq.~(\ref{eq:energy3}), and dividing it by $E_{\rm iso}$
\begin{equation}
    \frac{E^{obs}_{\rm P-GRB}}{E_{\rm iso}} =  \frac{a T^4_{obs}}{16\Gamma^2}\sigma_T \frac{{\cal B}}{m_N c^2} \Delta_{d},
\end{equation}
and the second equation is obtained from the energy conservation
\begin{equation}
1 = \frac{E^{obs}_{ \rm P-GRB}}{E_{\rm iso}} + \frac{E_{\rm Kinetic}}{E_{\rm iso}}\, 
\label{eq:energy6}
\end{equation}
where $E_{\rm Kinetic}$ is the kinetic energy of the baryonic PEMB pulses
\begin{equation}
E_{\rm Kinetic} = (\Gamma -1 ) M_B c^2\, .
\label{eq:energy7a}
\end{equation}
By substituting Eq.~(\ref{eq:energy7a}) in Eq.~(\ref{eq:energy6}) we have
    \begin{equation}\label{eq:energy7}
        {\cal B} = \frac{1}{\Gamma -1}\left(1-\frac{E^{obs}_{\rm P-GRB}}{E_{\rm iso}}\right).
    \end{equation}
    
%
%

\begin{figure*}
\centering
[a]\includegraphics[width=0.45\hsize,clip]{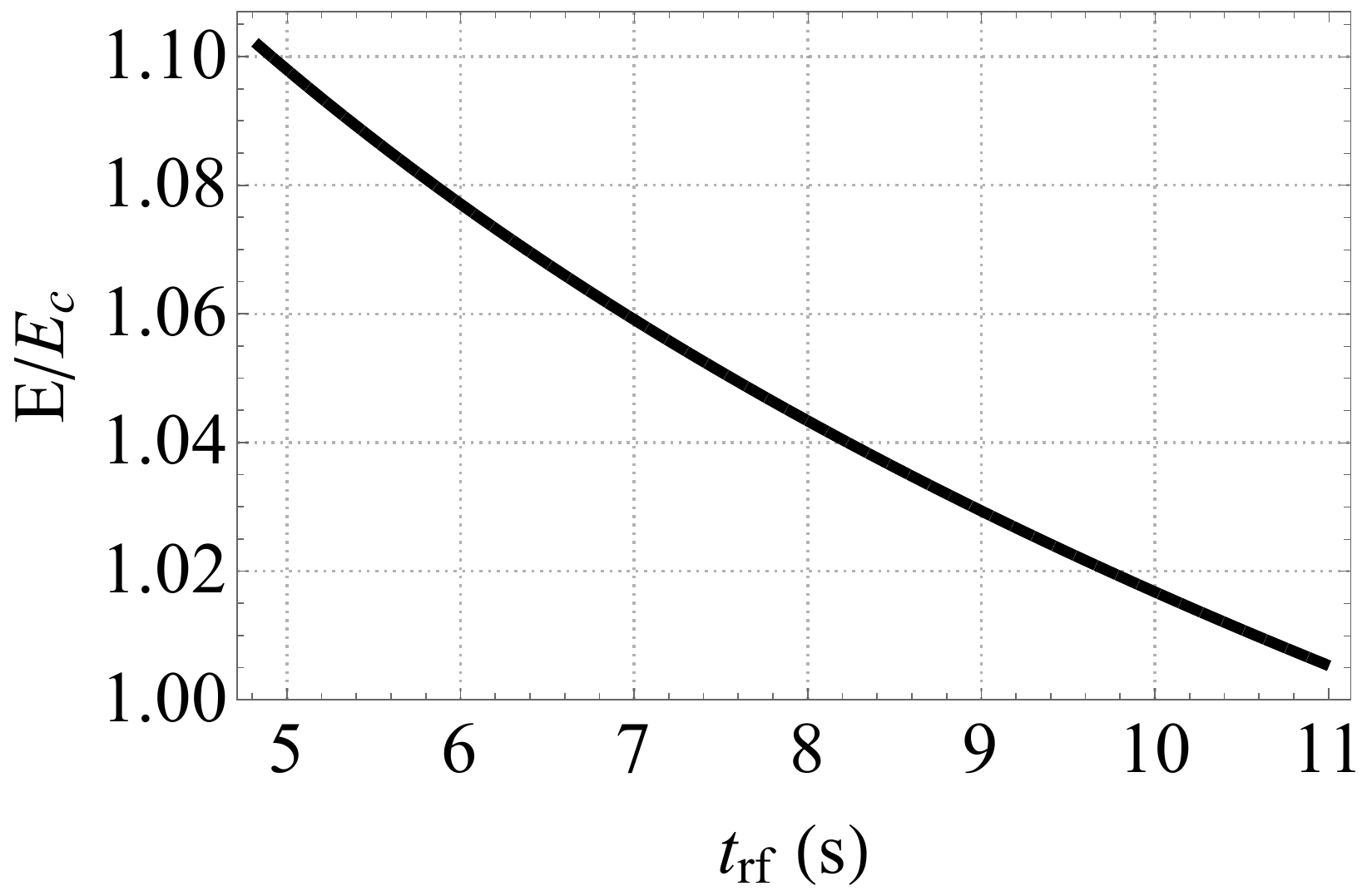}
[b]\includegraphics[width=0.47\hsize,clip]{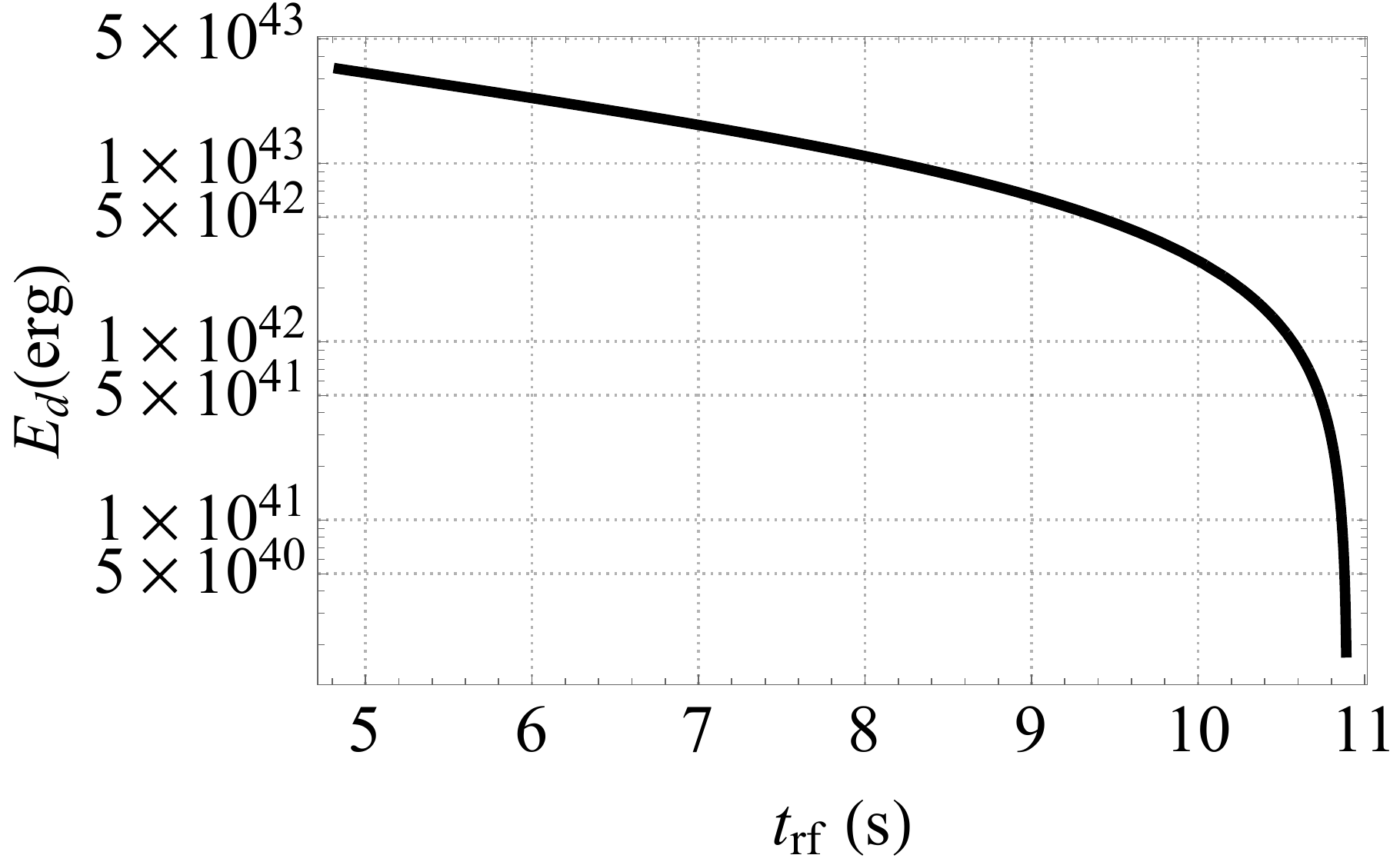}
[c]\includegraphics[width=0.46\hsize,clip]{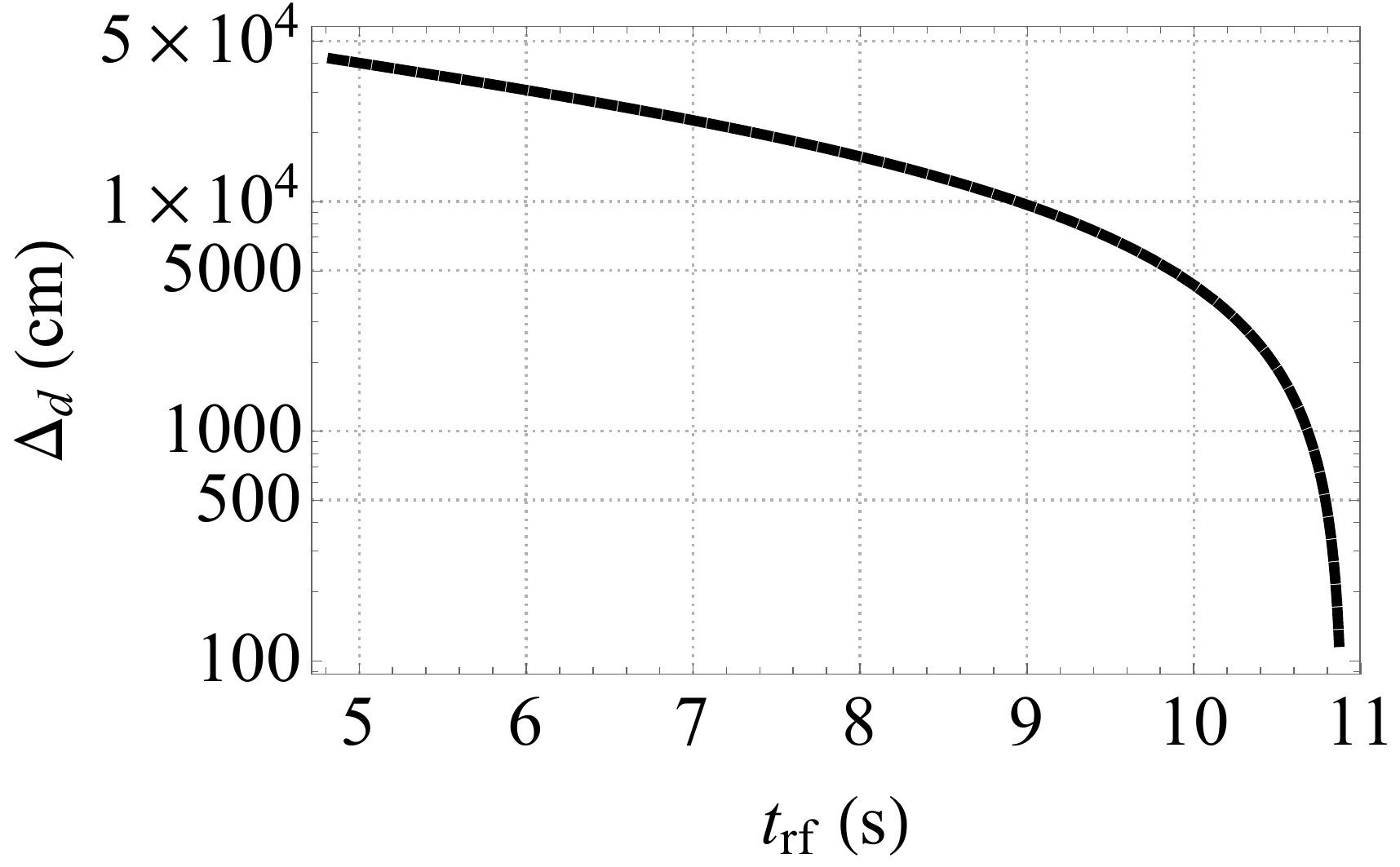}
[d]\includegraphics[width=0.45\hsize,clip]{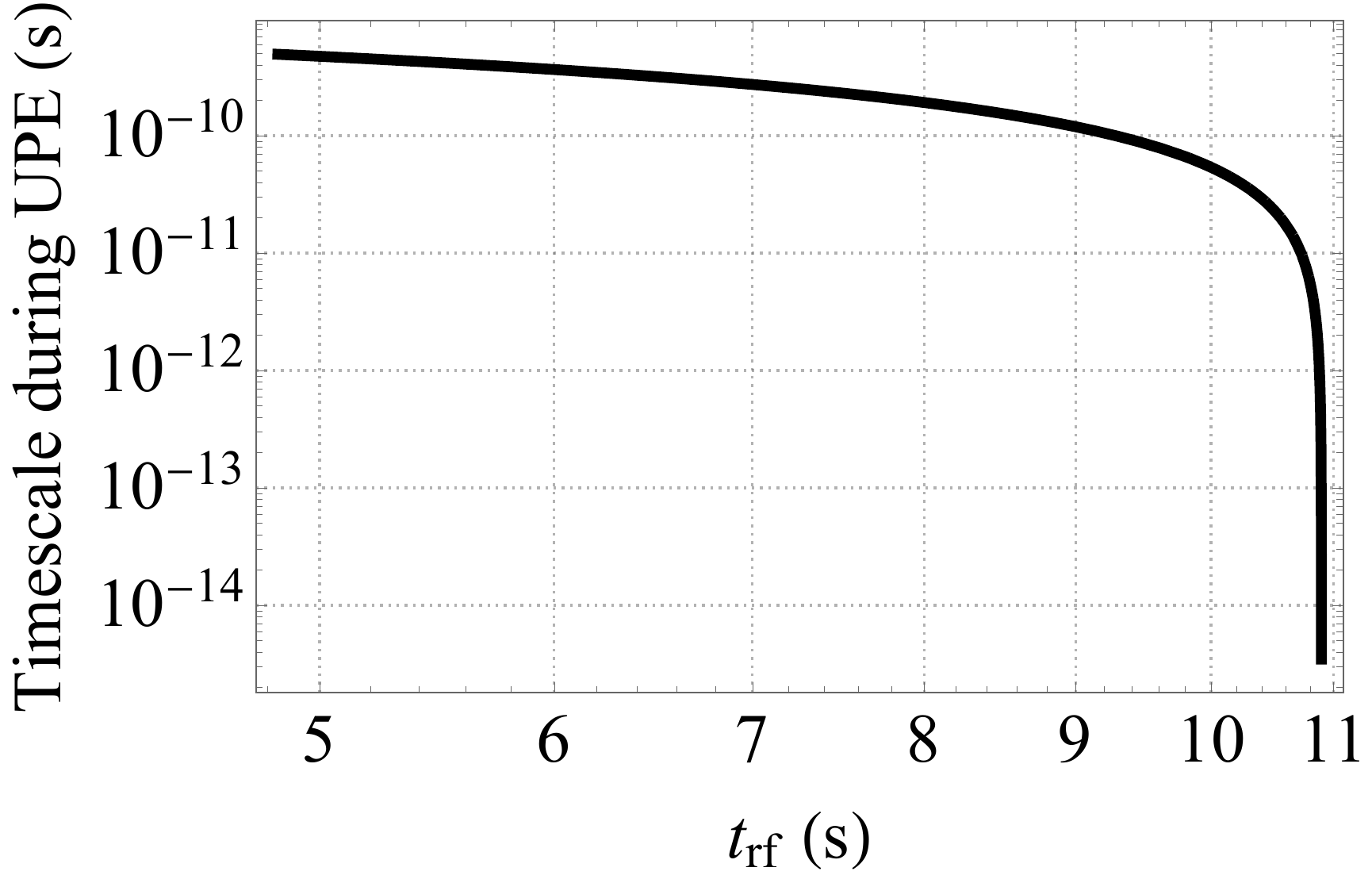}
[e]\includegraphics[width=0.45\hsize,clip]{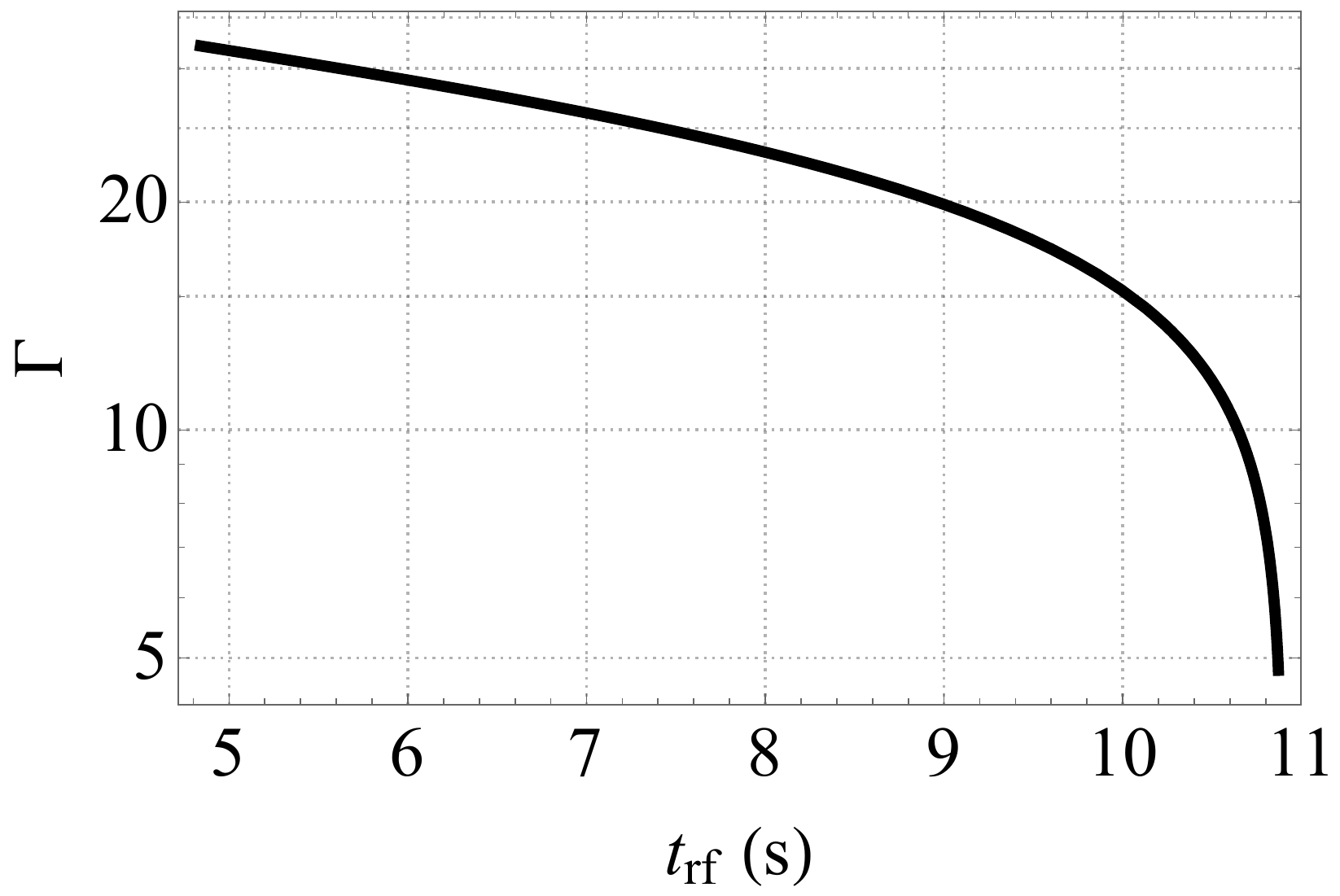}
[f]\includegraphics[width=0.46\hsize,clip]{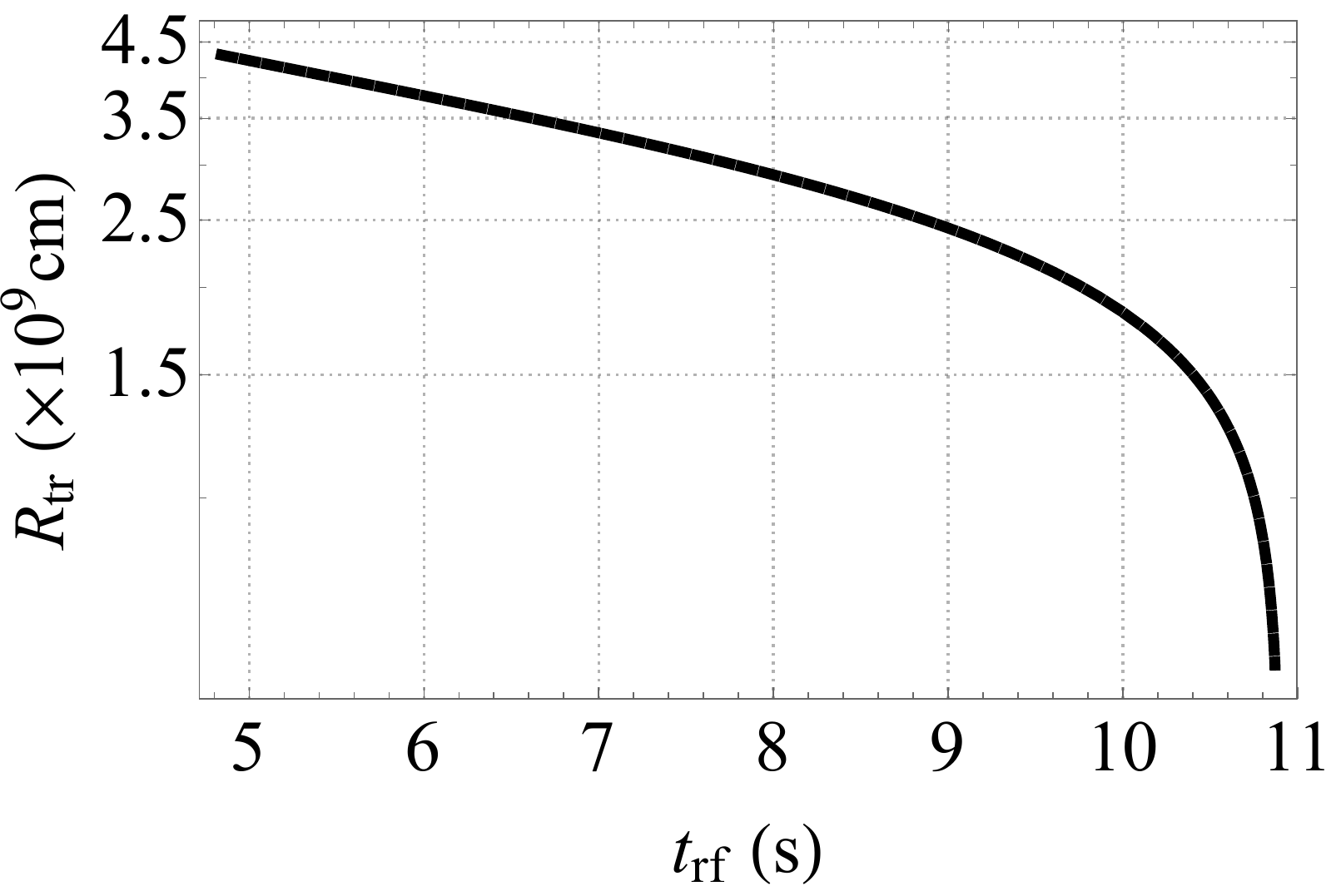}
\caption{ \textbf{[a]:} The evolution of the overcritical electric field  reaching its critical value at the end of UPE phase (t$_{\rm rf}=$10.89~s). \textbf{[b]:} The energy stored in dyadoregion obtained from Eq. (\ref{Eemxi}). \textbf{[c]:} The width of dyadoregion obtained from Eq. (\ref{eq:width}). \textbf{[d]:} Repetition time scale of the inner engine from Eq. (\ref{tauinner}). \textbf{[e]:} The Lorentz factor, $\Gamma$ which tends to unity at  t$_{\rm rf}=$10.89~s confirming the the end of UPE. \textbf{[f]:} The transparency radius; see Sec. \ref{sec:magnetic}.  All values are plotted as a function of rest-frame time for GRB 180720B during the UPE phase {($t_{\rm rf}=4.84$--$10.89$~s)}. }
\label{fig:eduringupe}
\end{figure*}

\section{Magnetic field and transparency condition and timescale of radiation during the UPE phase } \label{sec:magnetic}

The time evolution of the mass and spin of BH during the UPE was discussed in in Sec~\ref{sec:massupe}; see Fig.~\ref{massspinupe}. In order to calculate the magnetic field during the UPE phase, i.e., in the time interval  {$4.84<t_{\rm rf}< 10.89$~s,} we assume that the electric field therein is overcritical, which guarantees the occurrence of the UPE phase. Therefore, we infer a magnetic field strength $B_0=2.14 \times 10^{14}$~G such that the electric field given by Eq.~(\ref{eq:ER2}), at the end of the UPE phase at $t_{\rm rf}= 10.89$~s, fulfills $|E_{r_+}|= E_c$.

For this magnetic field, the dyadoregion energy at {$t_{\rm rf}=4.84$ is $5 \times 10^{43}$~erg} obtained from Eq.~(\ref{Eemxi}). Figures~\ref{fig:eduringupe}[a] and \ref{fig:eduringupe}[b] show the evolution of magnitude of electric field and the dyadoregion energy during the UPE phase. The total isotropic energy of the UPE phase is $E^{\rm UPE}_{\rm iso}=2.23 \times 10^{53}$~erg, consequently, there exist $\sim 10^{9}$ PEMB pulses during UPE phase. 

For the first PEMB pulse, assuming $B_0=1.87 \times 10^{14}$~G, the width of the dyadoregion at transparency point is {$\Delta_{\rm d}= 4.1 \times  10^{4}$~cm;} obtained from Eq.~(\ref{eq:width}). From the hierarchical structure of UPE phase in this GRB presented by Eq.~\ref{eq:ratio}, we have $E^{\rm obs}_{\rm P-GRB}/E_{\rm iso}\sim 0.03$ and the temperature $k T_{\rm obs}\sim 50$~keV; see Table.~\ref{tab:180720B}.

With these and following the previous section the transparency radius {$ R^{\rm tr} = 4.5 \times 10^9~\rm cm,$the baryon load parameter ${\cal B} = 3.1 \times 10^{-2},$} and finally the Lorentz factor $\Gamma = 38$, are obtained.

After the first PEMB pulse, whose energetics ($\Delta E$) is paid by the rotational energy of the BH (by reducing the $\Delta J$ from the angular momentum of the BH), the system starts over with the new value of the \textcolor{black}{BH mass, angular momentum and} effective charge\textcolor{black}{, as explained above}.

\begin{table}
\small\addtolength{\tabcolsep}{12pt}
\caption{ The parameters of the inner engine and the transparency point, obtained from the starting time of the UPE phase for GRB 180720B {($t_{\rm rf}=4.84$~s)} and GRB 190114C ($t_{\rm rf}=1.9$~s). \label{tab:comparision}}             
\centering  

\begin{tabular}{|c||c|c|c}       
\hline  
~&GRB180720B  &GRB190114C\\
~&  &\\
\hline  \hline 
$ R^{\rm tr}$ (cm)& $4.5\times 10^9$  & $9.4\times 10^9$ \\
\hline 
${\cal B}$& $3.1 \times 10^{-2}$ & $5.1 \times 10^{-3}$ \\
\hline 
$\Gamma $& 30 & 139 \\
\hline 
$\tau_{\rm q} $ (s)& $5 \times 10^{-10}$ & $3.1 \times 10^{-9}$ \\
\hline
$\Delta_d $ (cm) & $4.1 \times 10^{4}$ & $1.8 \times 10^{5}$\\
\hline
$B_0 $ (G) & $1.87 \times 10^{14}$ & $2.3 \times 10^{14}$\\
\hline

$|\mathbf{E}|/E_c $ & $1.11$ & $1.25$\\

\hline                                   
\end{tabular}

\end{table}

We infer from the MeV luminosity, the evolution of the radiation timescale $\tau_q(t)$  of the PEMB pulses by requiring it to explain the MeV emission energetics, i.e.:
\begin{equation}\label{tauinner}
  \tau_q(t)=\frac{E_{(r_+(t),r_d(t))}}{L_{\rm MeV}},
\end{equation}
where the $E_{(r_+(t),r_d(t))}$ is the energy of dyadoregion from Eq.~(\ref{Eemxi}), determined from the new values of $J$ and $M$ for each PEMB pulse, and $L_{\rm MeV}$ is the MeV luminosity obtained from the best fit in Sec.~\ref{sec:massupe}.

These parameters obtained from the starting time of the UPE phase, are similar to those of GRB 190114C; see Table~\ref{tab:comparision}. The evolution of the PEMB pulse timescale, the Lorentz $\Gamma$ factor, transparency radius, are shown in Fig.~\ref{fig:eduringupe} [d], Fig.~\ref{fig:eduringupe}[e] and Fig.~\ref{fig:eduringupe}[f], respectively.

{
We recall that the inner engine model has been first motivated to explain the GeV emission of GRBs as powered by an electrodynamical process that extracts the rotational energy of the newborn Kerr BH \cite{2019ApJ...883..191R, 2021A&A...649A..75M, 2022ApJ...929...56R}, and in \cite{2021PhRvD.104f3043M} for GRB 190114C and here for GRB 180720B it has been extended to explain the UPE phase. Following \cite{2021A&A...649A..75M, 2021PhRvD.104f3043M}, we summarize some key takeaways of our approach with respect to existing literature on this subject, in particular from numerical simulations.
}

{
There is a vast literature about magnetic fields around BHs and how they may act in a mechanism that could extract the mass-energy of a Kerr BH. \citet{1975PhRvD..12.2959R} made an early attempt using a matter-dominated magnetized plasma accreting in a disk around a pre-existing Kerr BH. They used the infinite conductivity condition, $F_{\alpha \beta} u^\beta = 0$, where $F_{\alpha \beta}$ is the electromagnetic field tensor and $u^\beta$ is the plasma four-velocity, leading to $\mathbf{E}\cdot \mathbf{B}=0$. Under these conditions, the acceleration of particles and processes of energy extraction were not possible. This work was further developed by \citet{1977MNRAS.179..433B}, who introduced the concept of gaps and spontaneous $e^+e^-$ pair creation in the context of a BH, closely following the pulsar theory by \citet{1971ApJ...164..529S} and \citet{1975ApJ...196...51R}, to have regions in the magnetosphere where $\mathbf{E}\cdot \mathbf{B} \neq 0$. They imposed a force-free condition, $F_{\alpha \beta} J^\beta = 0$, where $J^\beta$ is the current density. Their aim was to produce an ultrarelativistic matter-dominated plasma whose bulk kinetic energy could be used to explain the energetics of a jet at large distances from the BH. The alternative view of \citet{1982MNRAS.198..339T} extended the work of \citet{1977MNRAS.179..433B} and analyzed the problem of matter-dominated accretion in a magnetic field anchored to a rotating surrounding disk. The physical system, however, remained the same of \citet{1977MNRAS.179..433B}.
}

{
More recently, numerical simulations based on different models with respect to the one used in this article have been developed with the premise that the background electric field of a electro-vacuum solution (like the Papapetrou-Wald solution) might be screened from the surrounding plasma in the magnetosphere (see e.g. \citet{2005MNRAS.359..801K} and \citet{2019PhRvL.122c5101P}). These simulations have mainly addressed the physics of relativistic jets of plasma emerging from active galactic nuclei and x-ray binary systems and a especially detailed treatment and review of their theoretical models is presented by \citet{2005MNRAS.359..801K}. The choice of parameters and physical processes are different from the ones we have used for the GRB analysis. In our approach, we have been guided by the theoretical explanation of the following crucial observations of GRBs: (1) the time-resolved spectral analysis of the UPE phase; and (2) the MeV luminosity observed by Fermi-GBM. From this, we have identified the physical processes and parameters that have to be fulfilled in order to fit the vast amount of high-quality observational data. Their parameters enforce the condition $\mathbf{E} \cdot \mathbf{B} \neq 0$, while we use the Papapetrou-Wald solution which naturally possesses regions fulfilling such a condition in the BH vicinity.
}

{In our model, the magnetic field inherited from the collapsed NS is rooted in the surrounding material, and the electric field is created by the interaction of the gravitomagnetic field of the rotating BH with the external magnetic field. Since the electric field is assumed to be overcritical at the beginning, in a very short timescale of the order of the Compton time, $\hbar/(m_e c^2) \sim 10^{-21}$ s, which is much shorter than any electromagnetic process, it is originated a region dominated by the high density and high pressure of the neutral PEMB pulse. The PEMB pulse self-accelerates to the ultrarelativistic regime and finally reaches transparency at a radius $\sim 10^{10}$ cm.
}

{
As soon as the BH forms, the first and the most efficient process in action to produce the $e^+e^-$ plasma and, consequently decreasing the rotational energy of BH, occurs through the Schwinger critical field pair production. Since an overwhelming amount of pair plasma is created in quantum timescales, the plasma expansion by its internal pressure starts well before any electric field screening. This process takes a fraction of angular momentum of the Kerr BH. The BH then is left with a slightly smaller angular momentum $J^*= J - \Delta J$, with $\Delta J/J \sim 10^{-9}$, being $\Delta J$ the angular momentum extracted to the BH and the same magnetic field. This process leads to a new, lower value of the induced electric field. This process continues up to the moment when the electric field becomes undercritical.
}

{
The expanding $e^+e^-$ photon plasma sweeps matter in the cavity reducing the density of the latter to values as low as $\sim 10^{-14}$ g cm$^{-3}$, as shown by numerical simulations in \citet{2019ApJ...883..191R}. This low-density ionized plasma is needed to fulfill an acceleration of charged particles leading to the electrodynamical process around a newborn BH. This density is much lower the Goldreich-Julian density, for instance $\rho_{\rm GJ}\sim 10^{-11}$ g cm$^{-3}$, obtained for the present inner engine parameters. Moreover, the matter energy density inside the cavity is negligible comparing to the electromagnetic energy density, namely $\rho_M/(B^2 - E^2)\sim 10^{-14}$, while in \citet{2005MNRAS.359..801K} (see also \cite{2019PhRvL.122c5101P}), this ratio is $0.05$ or higher.
}

\section{Discussion and Conclusions}\label{sec:conc}


Following a new paradigm opened by the theoretical understanding and data analysis of GRB 190114C \cite{2019arXiv190404162R,PhysRevD.104.063043}, we have analyzed in this paper the UPE phase of GRB 180720B. We have here shown that also in GRB 180720B, a time-resolved spectral analysis conducted on shorter and shorter time intervals reveals the hierarchical structure of the UPE. Namely, the spectrum of the UPE phase, obtained in multiply rebinned time intervals, holds its features and is always fitted by a BB+CPL model (see Fig. \ref{alltogether}). We have shown the statistical significance of such a structure down to a time resolution of $0.11$ s.

We have then linked the above hierarchical structure of the UPE phase to a sequence of microphysical elementary events in the QED regime of the inner engine, occurring on a timescale of $\tau_q \sim 10^{-9}$~s. The understanding of the underlying quantum nature is not possible without the discovery of the observed hierarchical structure of the UPE phase.

The inner engine is composed of a Kerr BH rotating in a uniform magnetic field $B_0$, aligned with the BH rotation axis, described by the Papapetrou-Wald solution, immersed in a rarefied plasma. The gravitomagnetic interaction of the rotating BH with the magnetic field induces an electric field. The process that originates the $10$~keV--$10$~MeV radiation is triggered by the vacuum polarization that occurs when the induced electric field in the inner engine is overcritical, i.e., $|{\bf E}|>E_c$. This process forms around the BH an optically thick pair $e^+~e^-~\gamma$ plasma whose high internal pressure drives its self-accelerating expansion. During the expansion, the plasma is loaded with baryons forming the PEMB pulse that reaches ultrarelativistic regime with $\Gamma\sim 30$ and the transparency point where the radiation becomes observable \cite{2010PhR...487....1R}. We assume that the magnetic field $B_0 \sim10^{14}$~G is constant during the UPE phase.

In the radiation timescale of the PEMB pulses, $\tau_q \sim 10^{-9}$~s, the above process extracts $\Delta J\sim  10^{-9}~J$ of angular momentum of the Kerr BH, leaving it with a new, lower angular momentum $J^* = J- \Delta J$. Since the magnetic field is assumed constant during the UPE phase, the new value of the induced electric field is lower. Then, the system starts a new vacuum polarization process in the presence of the same magnetic field $B_0$, and a new effective charge of $Q^*_{\rm eff}= Q_{\rm eff}-\Delta Q_{\rm eff}$, where $\Delta Q_{\rm eff} =2 B_0 \Delta J$. It leads to the production of approximately $10^9$ PEMB pulses, which one after another reach the transparency point and their radiations form the UPE phase. This process continues till the electric field lowers to $|{\bf E}|<E_c$.

The magnetic field in this scenario is inherited from the NS and is amplified in the gravitational collapse to a BH. Consequently, the electric field and consequent effective charge, $Q_{\rm eff}=2 B_0 J G/c^3$, are induced by the gravitomagnetic interaction of the rotating BH with the external magnetic field \citep{2020ApJ...893..148R, 2022ApJ...929...56R}. The electric field is overcritical during the UPE phase. In a quantum timescale, $\hbar/(m_e c^2) \approx 10^{-21}$ s, the dyadoregion characterized by the high density and high pressure of the $e^+e^-\gamma$ plasma develops and dominates over any other electromagnetic process \cite{2010PhR...487....1R}.

\begin{acknowledgements}
\textcolor{black}{We thank the Referee for the deep remarks and comments which have improved the presentation of the results and readability of the paper.}
\end{acknowledgements}


\end{document}